\documentclass[apj,useAMS,usenatbib,usegraphicx,upmath]{emulateapj} \usepackage{amsmath} \usepackage{amsfonts} \usepackage{amssymb}  \usepackage{apjfonts}\usepackage{color}

\pdfoutput = 1

\shorttitle{The thermosphere of HD209458b}
\shortauthors{T.T. Koskinen et al.}

\begin{document}

\title{Characterizing the thermosphere of HD209458b with UV transit observations}

\author{T. T. Koskinen\altaffilmark{1,2}, R. V. Yelle\altaffilmark{1}, P. Lavvas\altaffilmark{1}, N. K. Lewis\altaffilmark{1}}
\altaffiltext{1}{Lunar and Planetary Laboratory, University of Arizona, 1629 E. University Blvd., Tucson, AZ 85721--0092; tommi@lpl.arizona.edu}
\altaffiltext{2}{Honorary Fellow, Department of Physics and Astronomy, University College London, Gower Street, London WC1E 6BT, UK}

\begin{abstract}
Transmission spectroscopy at UV wavelengths is a rich and largely unexplored source of information about the upper atmospheres of extrasolar planets.  So far, UV transit observations have led to the detection of atomic hydrogen, oxygen and ionized carbon in the upper atmosphere of HD209458b.  The interpretation of these observations is controversial -- it is not clear if the absorption arises from an escaping atmosphere interacting with the stellar radiation and stellar wind, or the thermosphere inside the Roche lobe.  In this paper, we introduce an empirical model that can be used to analyze UV transit depths of extrasolar planets.  We use this model to interpret the transits of HD209458b in the H~Lyman~$\alpha$ and the 1304~\AA~O~I triplet emission lines.  The results indicate that the mean temperature of the thermosphere is $T =$~8,000--11,000~K and that the H$_2$/H dissociation front is located at pressures between $p =$~0.1--1~$\mu$bar, which correspond to an altitude of $z \approx$~1.1~R$_p$.  The upper boundary of the model thermosphere is located at altitudes between $z =$~2.7--3~R$_p$, above which the atmosphere is mostly ionized.  We find that the H~I transit depth reflects the optical depth of the thermosphere in the wings of the H~Lyman~$\alpha$ line but that the atmosphere also overflows the Roche lobe.  By assuming a solar mixing ratio of oxygen, we obtain an O~I transit depth that is statistically consistent with the observations.  An O~I transit depth comparable to or slightly larger than the H~I transit depth is possible if the atmosphere is undergoing fast hydrodynamic escape, the O/H ratio is supersolar, or if a significant quantity of neutral oxygen is found outside the Roche lobe.  We find that the observations can be explained solely by absorption in the upper atmosphere and extended clouds of ENAs or atoms strongly perturbed by radiation pressure are not required.  Due to the large uncertainty in the data, repeated observations are necessary to better constrain the O~I transit depths and thus the composition of the thermosphere.                   
\end{abstract}


\section{Introduction}
\label{sc:intro}

Visible and near-IR spectroscopy of transiting extrasolar planets has led to the detection of Na~I, H$_2$O, CH$_4$, CO, and CO$_2$ \citep{charbonneau02,tinetti07,swain08,swain09} in the atmospheres of extrasolar giant planets (EGPs).  Discoveries such as these have helped to shift the emphasis in the study of extrasolar planets from mere detection towards characterization of the dynamics and composition of their atmospheres.  Modeling and observations of a large variety of objects allow us to constrain theories of the formation and evolution of different planets and their atmospheres.  In this respect, studies of mass loss from the atmospheres of close-in EGPs and terrestrial planets are of particular interest.  In addition to being interesting in their own right, these studies are valuable for constraining models of planetary atmospheres during the early history of the solar system.   

Approximately 25\% of the currently known extrasolar planets orbit within 0.1 AU from their host stars\footnote{See the Extrasolar Planets Encyclopaedia, maintained by J. Schneider, at http://www.exoplanet.eu/.}.  At such small orbital distances many of these planets are subject to intense stellar irradiation.  In particular, the outermost layers of the atmosphere are strongly ionized and heated by the absorption of FUV and EUV radiation, which leads to the formation of the thermosphere \citep[e.g.,][]{yelle04,garciamunoz07,koskinen07}.  Modeling indicates that the thermospheres of close-in extrasolar giant planets are hot, greatly extended, and their composition is dominated by atomic constituents.  The same should apply to close-in super-Earths orbiting M dwarf stars.   These are the closest types of object to an Earth-like habitable planet whose atmospheres we may be able to probe with space-born or even ground-based telescopes in the near future \citep[e.g.,][]{palle09,charbonneau09}.  As a consequence, the UV transit depths of extrasolar planets should be much larger than the visible and IR transit depths because the observations probe the upper atmosphere that strongly absorbs UV radiation in electronic resonance lines.

Transit observations of extrasolar planets in the FUV were pioneered by \citet{vidalmadjar03} (hereafter VM03) and \citet{vidalmadjar04} (hereafter VM04) who detected atomic hydrogen (H~I), oxygen (O~I), and ionized carbon (C~II) in the upper atmosphere of the transiting planet HD209458b.  All of these detections were obtained by using the Space Telescope Imaging Spectrograph (STIS) onboard the Hubble Space Telescope (HST).  VM03 used the G140M grating to observe three transits of HD209458b in the wavelength range covering the stellar H~Lyman~$\alpha$ (H~Ly$\alpha$) emission line.  The spectral resolution of these observations was $\sim$0.08~\AA, which allowed for the details of the line profile to be resolved.  VM04 used the G140L grating with a low spectral resolution of $\sim$2.5~\AA~to observe four transits in the wavelength range of [1180,1710]~\AA~and detected absorption in the O~I and C~II stellar emission lines. 

VM03 deduced a H~I transit depth of 15~$\pm$~4~\% from the ratio of the flux in two wavelength regions around the core of the H~Ly$\alpha$ line to the flux in the wings of the line during transit.  Based on this observation, they concluded that H~I overflows the Roche lobe with a mass loss rate of $dM/dt >$~10$^{7}$~kg~s$^{-1}$.  They also suggested that the planet is followed by a cometary tail that is shaped by stellar radiation pressure acting on the escaping hydrogen.  Later, \citet{benjaffel07} (hereafter BJ07) and \citet{benjaffel08} (hereafter BJ08) presented a thorough and convincing reanalysis of the same data.  Based on this analysis, BJ08 argued that there is no evidence for a cometary tail in the transit light curve or the detailed transit depth measurements.  Further, he showed that the observed absorption may arise from atomic hydrogen below the Roche lobe and that the H~I absorption line profile is broadened by thermal and natural broadening in the thermosphere of the planet.  However, his results also imply that hydrogen overflows the Roche lobe, and thus the atmosphere is still evaporating \citep{vidalmadjar08}.        

VM04 used the low resolution G140L data to obtain O~I and C~II transit depths of 13~$\pm$~4.5 and 7.5~$\pm$~3.5~\%, respectively.  They also calculated a full-width H~I transit depth of 5~$\pm$~2~\% from the unresolved H~Ly$\alpha$ line and claimed that this transit depth is consistent with the much stronger absorption observed within a limited section of the line profile by VM03.  Further, they argued that the large O~I and C~II transit depths are possible because these species overflow the Roche lobe and the absorption lines are broadened by the velocity dispersion of the escaping gas.  Thus the observations were interpreted as proof that the atmosphere is undergoing fast hydrodynamic escape.    

\citet{benjaffel10} (hereafter BJ10) published a thorough reanalysis of the G140L data.  They obtained revised full-width H~I, O~I, and C~II transit depths of 6.6~$\pm$~2.3~\%, 10.5~$\pm$~4.4~\%, and 7.4~$\pm$~ 4.7~\%, respectively.  We note that these depths are only 2, 1.93, and 1.15$\sigma$, respectively, away from the FUV continuum transit depth of $\sim$2~\%.  The more detailed H~I transit depth measurements reported by BJ08 provide stronger constraints for the H~Ly$\alpha$ line but no such constraints are available for the O~I and C~II transits.  In order to explain the large transit depths in these lines, BJ10 argued that oxygen and ionized carbon are preferentially heated to a temperature more than ten times higher than the temperature of hydrogen within a layer in the atmosphere located between $\sim$2.25~R$_p$ and the boundary of the Roche lobe at 2.9~R$_p$.  

Most analyses of the HD209458b UV absorption signatures to date have either been limited to first order deductions, such as the effective size of the absorbing obstacle (VM03,VM04) or been based on complicated first principle models for the atmosphere \citep[][]{tian05,garciamunoz07,murrayclay09,benjaffel10}.  We believe that there is an important role for an intermediate class of models that satisfy some basic physical constraints, but parameterize aspects of the atmosphere that are difficult or impossible to predict accurately.  For example, it is well established that the thermosphere of HD209458b should be composed primarily of H and H$^+$, but the location of the transition from H$_2$ to H is uncertain with different physical models making vastly different predictions \citep{liang03,yelle04,garciamunoz07}.  Moreover, the boundary between the atmosphere and interplanetary space is dependent upon the unknown strength of the stellar wind and the planetary and interplanetary magnetic field and has yet to be modeled in a realistic fashion.  Finally, although it seems well established that the temperature is of the order 10,000~K, the precise value depends on the heating efficiencies, which have yet to be calculated, and radiative cooling by minor species, which is not included in any of the models.  Because of these uncertainties it is important to analyze the data in a way that makes clear what aspects of the atmosphere are constrained by the observations and which are not.                

In this paper we introduce a generic methodology that can be used to interpret UV transit light curves in stellar emission lines.  We demonstrate this methodology by using a simple empirical model of the thermosphere to analyze the H~I and O~I transit depths of HD209458b summarized by BJ10.  In Section~\ref{sc:methods} we introduce a model for calculating transit light curves for planets with extended atmospheres and discuss the basic features of the model thermosphere.  In Section~\ref{sc:results} we discuss the H~I transit depth measurements in detail and confirm that they \textit{can} be explained by absorption by atomic hydrogen below the Roche lobe.  Nevertheless, we also show that the core of the H~Ly$\alpha$ absorption line is optically thick up to the Roche lobe and that the atmosphere is evaporating.  Further, we demonstrate that the disagreement between BJ08 and VM03 is due to differences in the treatment of the data and different definitions of the transit depth.       

In the rest of Section~\ref{sc:results}, we discuss the transits in the O~I triplet lines and show that the empirical model thermosphere with a solar abundance of oxygen can be used to obtain transit depths that are statistically consistent with the observations.  We also address the feasibility of the idea that energetic oxygen atoms are present in the thermosphere of HD209458b and present alternative ways to explain O~I transit depths that are comparable to or larger than the full-width H~I depth.  We conclude Section~\ref{sc:results} by discussing a variety of different models for the thermosphere of HD209458b and use them to calculate transit depths.  In Section~\ref{sc:discussion} we discuss the feasibility of our assumptions and in Section~\ref{sc:conclusion} we summarize our findings and conclusions.              

\section{Method}
\label{sc:methods}

\subsection{Transit light curves}
\label{subsc:transits}

In order to compare the observed UV transit depths with models of the thermosphere, we need to calculate the flux decrement of the stellar emission lines observed at Earth at different times $t$ during the transit.  The wavelength-integrated transit depth, $T_{\delta \lambda} \left( t \right)$, is given by:
\begin{eqnarray}
T_{\delta \lambda} \left( t \right) &=&  \frac{1}{d_E^2 F_o} \int_{\delta \lambda} S (\lambda - \lambda') d\lambda \nonumber \\ 
&\times& \int_{A_d} I \left(\lambda, \mathbf{x} \right) \exp \lbrack - \tau ( \lambda, t , \mathbf{x} ) - \tau_{ISM} (\lambda) \rbrack dA_d
\label{eq:integ_trans}
\end{eqnarray}
where $d_E$ is the distance to Earth, $S (\lambda - \lambda')$ is the instrument response function, $\delta \lambda$ corresponds to the spectral resolution of the instrument, $I \left( \lambda,\mathbf{x} \right)$ is the specific intensity of the star expressed as a function of position $\mathbf{x}$ on the stellar disk of area $A_d$, $\tau \left( \lambda, t, \mathbf{x} \right)$ is the optical depth due to the planet and its atmosphere along the line of sight (LOS) from the host star to the observer, $\tau_{ISM} (\lambda)$ is the optical depth of the interstellar medium (ISM) along the sightline to the star, and $F_o$ is the out-of-transit flux within $\delta \lambda$ observed at Earth.  We note that interstellar absorption does not affect the transit depth unless the ISM is optically thick.  In that case the transmitted flux is so small that the observations cannot be used to calculate transit depths.  This affects the analysis of transit depths calculated from unresolved or partly resolved stellar emission lines. 

The optical depth $\tau \left( \lambda, t , \mathbf{x} \right)$ is a complicated function of time and position, and in general equation~(\ref{eq:integ_trans}) cannot be integrated analytically.  In order to integrate the equation numerically, we adopted a Cartesian coordinate system with the origin at the center of the stellar disk.  At time $t$, the coordinates of the center of the planet in this system are:
\begin{eqnarray}
d \left( t \right) &=& \xi \left( t \right) \sqrt{ 1 - cos^2 \lbrack \alpha \left( t \right) \rbrack} \\
b \left( t \right) &=& \xi \left( t \right) \cos \left( i \right) \cos \lbrack \alpha \left( t \right) \rbrack
\end{eqnarray}
where $d$ and $b$ are the absolute values of the $x$ and $y$ coordinates of the planet, respectively, $\xi$ is the distance between the centers of the star and the planet, $i$ is the inclination, and $\alpha$ is the angle between the LOS to Earth and the line joining the centers of the planet and the star \textit{in the orbital plane}.  The $y$ coordinate of the planet at the center of the transit, $b_c = \xi_c \cos \left( i \right) / R_*$, expressed in terms of the stellar radius $R_*$, is known as the \textit{impact parameter}.  The angle $\alpha$ is related to the true anomaly $\theta(t)$ of a given orbital position by $\alpha =  | \theta_c - \theta \left( t \right) |$ where $\theta_c$ is the true anomaly at the center of the transit.  If the orbit is circular, obtaining $\alpha$ is straightforward.  For an eccentric orbit the distance $\xi$ and true anomaly $\theta$ as a function of time can be calculated by solving Kepler's equation \citep[e.g.][]{koskinen09}.

\subsection{Optical depth}
\label{subsc:optical_depth}

Assuming that the atmosphere is spherically symmetric, the optical depth $\tau (\nu,z)$ at frequency $\nu$ along a LOS with a tangent altitude $z$ from the center of the planet is given by:
\begin{equation}
\tau \left( \nu,z \right) =  2 \int_{z}^{\infty} \frac{ \sigma_{\nu} (p,T) n_s \left( r \right) r dr}{\sqrt{ r^2 - z^2 }}
\label{eq:tau}
\end{equation}
where $n_s$ is the number density of the absorbing species $s$, $\sigma_{\nu}$ is the absorption cross section, $p$ is pressure, $T$ is temperature, and $r$ is the distance from the center of the planet.  We note that $p$ and $T$ depend on the altitude $r$ and thus $\sigma_{\nu}$ is a function of altitude.  For individual absorption lines the cross section is given by:
\begin{equation}
\sigma_{\nu} \left( p,T \right) = \frac{e^2}{4 \epsilon_o m_e c} \frac{f_o}{\sqrt{\pi} \Delta \nu_D} \phi_{\nu} \left( a,\upsilon \right)
\label{eq:sigma0}
\end{equation} 
where $f_o$ is the oscillator strength and $\phi_{\nu} \left( a,\upsilon \right)$ is the Voigt function.  The arguments of the Voigt function are given by:
\begin{equation*}
a = \frac{b_L}{2 \Delta \nu_D} \ \ \ \ , \ \ \ \ \upsilon = \frac{\nu - \nu_o}{\Delta \nu_D} \ \ \ \ , \ \ \ \ \Delta \nu_D = \frac{\nu_o}{c} \sqrt{ \frac{2 k T}{m_s} }
\end{equation*}
where $b_L$ is due to natural broadening, $\nu_o$ is the laboratory frequency of the absorption line, $m_s$ is the mass of the absorbing species, and $\Delta \nu_D$ is the Doppler broadening parameter.  

The relationship between the tangent altitude $z$ and the $x$ and $y$ coordinates of a position on the stellar disk is given by:  
\begin{equation}
z^2 \left( t,x,y \right) = \lbrack x - d( t ) \rbrack^2 + \lbrack y - b( t ) \rbrack^2
\end{equation}
This relationship can be used to calculate $z$ for any point on the stellar disk and the LOS optical depth $\tau (\nu,z)$ can then be calculated by integrating equation~(\ref{eq:tau}).  If the atmosphere is not spherically symmetric, the problem of obtaining the LOS optical depth as a function of position on the stellar disk is much more complicated.  In that case, equation~(\ref{eq:tau}) must be integrated separately for each latitude point, taking into account the longitudinal variation of the atmosphere along the LOS.  This technique may be relevant if, for instance, future observations of UV transit light curves indicate that the upper atmosphere deviates significantly from a spherical shape.    

As stated above, we integrated equation~(\ref{eq:integ_trans}) numerically in Cartesian coordinates centered on the stellar disk.  We used Simpson's rule to set up a sequence of one-dimensional integrals in the $y$ direction that cover the disk of the star in the $x$ direction.  This technique is iterative and the algorithm increases the number of points on the stellar disk until a certain predetermined numerical precision is achieved [see Section 4.6 of \citet{press92}].  The transit depths in the UV are of the order of 1--10~\%, and thus we required numerical errors less than 0.1~\% in all of our simulations.  Given a model atmosphere, equation~(\ref{eq:tau}) was integrated by using the method described by  \citet{smith90} (see Section~\ref{subsc:model_atmosphere} below).  The optical depths were calculated for a fixed altitude grid of the model atmosphere, and the values were interpolated to the appropriate stellar disk coordinates by the iterative algorithm that was used to integrate equation~(\ref{eq:integ_trans}).  Voigt functions were evaluated numerically by using the method of \citet{humlicek82}.                   

\subsection{Model atmosphere}
\label{subsc:model_atmosphere}

\begin{figure}
  \epsscale{1.15}
  \plotone{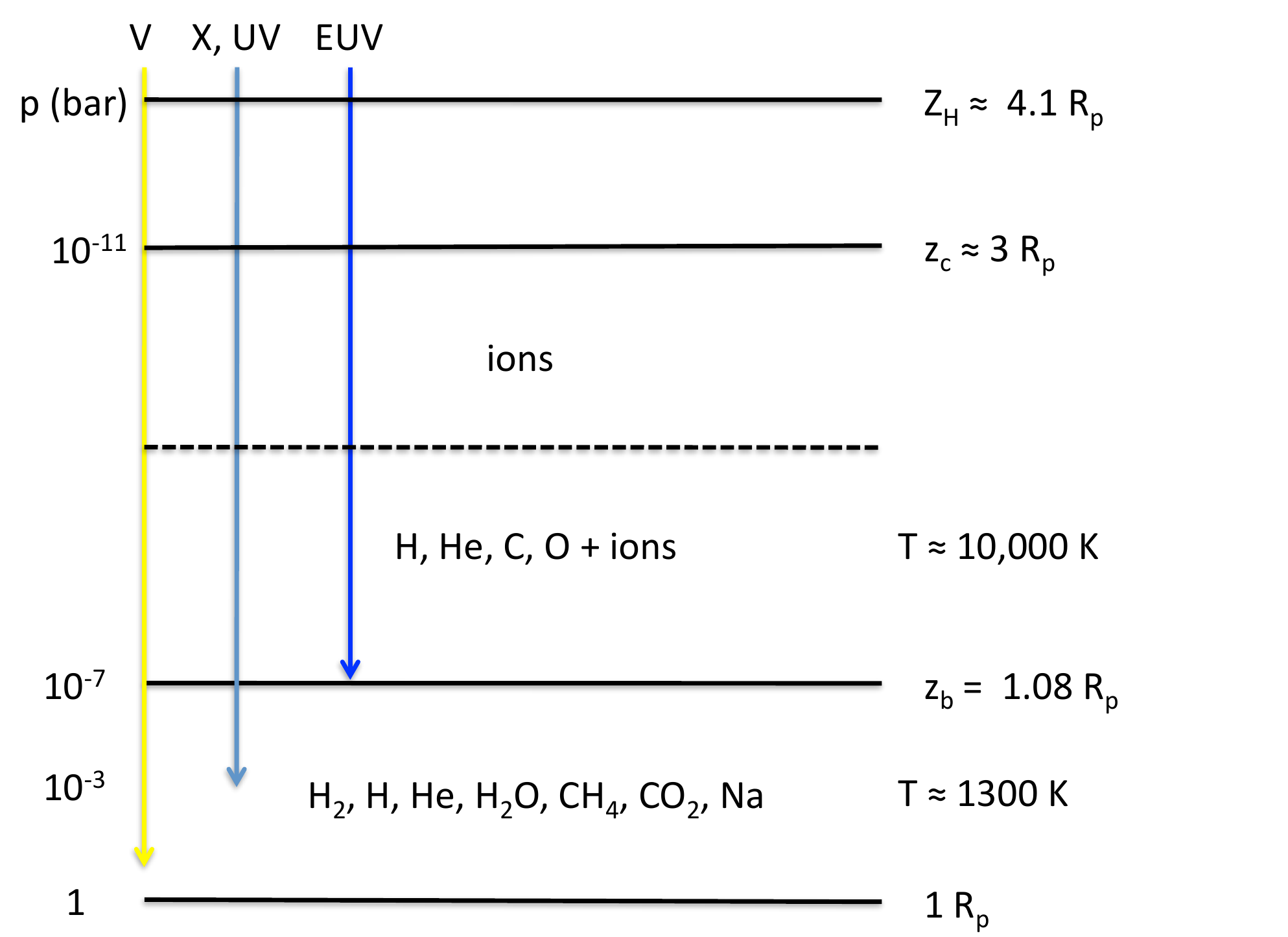}
  \caption{An empirical model of the atmosphere of HD209458b.  Several important transition altitudes are shown.  Above $z_b =$~1.08~R$_p$ ($p =$~10$^{-7}$~bar) molecules are dissociated, and atomic species of H, C, and O have been detected (VM03,VM04).  The H$^+$/H ionization front is located between $Z_e =$~2--5~R$_p$ and the ionization fronts of the other species may also be located at a similar altitude with H.  The critical altitude where the outflow velocity becomes supersonic in fast hydrodynamic escape is located at $z_c \approx$~3~R$_p$.  The Hill radius is $Z_{H} \approx$~4.1~R$_p$ and the distance of the Roche lobe from the center of the planet in the direction perpendicular to the LOS is $Z_{H \perp} \approx$~2.9~R$_p$ \citep{gu03}.  The vertical arrows show the different types of stellar radiation and their approximate penetration depths.  In addition to hydrogen and helium the list of atmospheric constituents in the lower atmosphere includes the species that have been detected in the atmosphere of HD209458b \citep[e.g.][]{charbonneau02,knutson08,swain09}.}
  \label{fig:model_atmos}
\end{figure}

Model atmospheres based on hydrodynamics and photochemistry are complicated and time-consuming to use.  In addition, they are based on many uncertain assumptions that cannot always be changed easily to match with the observations.  In order to fit the data without the bias due to any individual model, we have constructed a simple empirical model of the upper atmosphere of HD209458b.  This model is based on only a few free parameters that can be constrained by the generic features of the more complex models.  Figure~\ref{fig:model_atmos} shows the structure of the upper atmosphere and highlights some of the important transition altitudes of the model.  The lower boundary of the thermosphere is located at the pressure of $p_b =$~0.1~$\mu$bar, at an altitude $z_b \approx$~1.1~R$_p$.  This choice is motivated by the fact that the stellar EUV radiation, which causes intense heating and ionization of the upper atmosphere, is almost fully absorbed above the 1~$\mu$bar level and as a result the volume heating rate peaks near the 0.1~$\mu$bar level \citep{koskinen10a}.  At higher altitudes H$_2$ is dissociated thermally and by dissociative photoionization.  This is also true of the more complex molecules, and at $z >$~1.1~R$_p$ the atmospheric constituents appear as atoms or atomic ions  \citep{yelle04,garciamunoz07}.  

We assume that the lower atmosphere below $z_b$ is opaque to the FUV radiation considered in this study, and that the mean effective temperature is $T =$~1300~K up to the 0.1~$\mu$bar level.  This temperature, together with the assumption of hydrostatic equilibrium, is used to constrain the vertical extent of the lower atmosphere.  We note that X rays and mid-UV radiation penetrate to the mbar levels, while the bulk of the visible radiation penetrates down to the $\sim$1~bar level or slightly above it.  Thus these types of radiation do not have a significant effect on the thermosphere where the conditions are largely determined by EUV heating and ionization.       

Several different models suggest that the temperature of the thermosphere is $T \sim$~10,000~K \citep[e.g.,][]{yelle04,garciamunoz07,murrayclay09,koskinen10a}.  Consequently, the pressure scale height is relatively large, reaching values of $H_p > 7000$~km.  This implies that the thermal escape parameter $\lambda_c = r / H_p <$~13, and thus several authors have suggested that the upper atmosphere of HD209458b is undergoing fast hydrodynamic escape\footnote{By fast hydrodynamic escape, we mean outflow that reaches supersonic velocity at some critical altitude.} \citep[e.g.][]{lammer03,tian05,garciamunoz07,murrayclay09}.  Despite this possibility we assume that the density profile is hydrostatic.  This assumption is justified because the density profile is approximately hydrostatic up to the critical level at $z_c \approx$~3~R$_p$ even when the atmosphere is escaping at supersonic speed at higher altitudes.

Based on the ratio of the vertical advection term to the pressure gradient term in the momentum equation, the outflow velocity must be nearly equal to or faster than the speed of sound for the departure from hydrostatic equilibrium to be significant globally.  Most models indicate that the mass loss rate from the atmosphere of HD209458b is limited to $dM/dt =$~10$^{7}$--10$^{8}$~kg~s$^{-1}$ by the available EUV energy.  The pressure at $z_c = $~3~R$_p$ in the empirical model shown in Figure~\ref{fig:model_atmos} is $p \approx$~0.04~nbar, which implies that the vertical velocity at the critical level is $v_r =$~1--10~km~s$^{-1}$.  Due to the conservation of mass, the vertical velocity is inversely proportional to density and thus it decreases steeply with decreasing altitude.  For example, at $p =$~1~nbar, which corresponds to the altitude of $z =$~1.7~R$_p$, the vertical velocity required to support mass loss is only $v_r =$~30--300~m~s$^{-1}$.  Such a slow vertical velocity does not cause a significant deviation from hydrostatic equilibrium in the thermosphere \citep[e.g.,][]{yelle04}.
    
Given a hydrostatic density profile, we integrated equation~(\ref{eq:tau}) numerically.  For the purposes of the numerical solution, we created a grid with an altitude spacing of 0.01~H$_p$ for the thermosphere.  The number density $n_s$ of species $s$ at altitude $r_k$ corresponding to the grid point $k$ is simply given by:   
\begin{equation}
n_s \left( r_k \right) = x_s n \left( r_{k-1} \right) \exp \lbrack \lambda_T \left( r_k \right) - \lambda_T \left( r_{k-1} \right) \rbrack
\label{eq:number_density}
\end{equation}
where $x_s$ is the volume mixing ratio of species $s$, and $n$ is the total number density.   The thermal escape parameter, $\lambda_T$, is given by:
\begin{equation}
\lambda_T \left( r_k \right) = \frac{G M_p m_{k-1}}{k T_{k-1} r_k}
\label{eq:lambda}
\end{equation}
where $m_{k-1}$ and $T_{k-1}$ are the mean molecular weight and temperature, respectively, at a lower grid point $k-1$.  

In order to match the observations we treated the number density at the lower boundary, $n_b$, the mean temperature of the thermosphere, $T$, and the mixing ratios $x_s$ as free parameters.  In practice, we used $p_b$ and $T$ to constrain $n_b$, and assumed a multiple of the solar composition to fix $x_s$.  The resulting mixing ratios of H, He, C, O, and N were used to calculate the mean molecular weight $m$.  The upper boundary of the model is at an altitude $Z_e$ above which the atmosphere is mostly ionized.  The altitude $Z_e$ need not be the same for different species and generally it depends on the ionizing flux, photoabsorption cross sections, photochemical reaction rates, and the outflow velocity.  We assumed that the the column densities of the neutral species are negligible above $Z_e$ and treated it as a free parameter for each species.        

We also assumed uniform mixing throughout the model atmosphere and that diffusive separation does not take place in the thermosphere.  This assumption can be justified because the large pressure scale height in the atmosphere of HD209458b may lead to efficient turbulent mixing and implies that the exobase is located at a relatively high altitude.  In general, diffusive separation occurs when the coefficient of turbulent (eddy) diffusion, $K_{\tau}$, is smaller than the coefficients of molecular diffusion, $D_s$, for different species.  The coefficient of eddy diffusion can be estimated crudely as $K_{\tau} \sim v_{\tau} H_p$ \citep[e.g.,][]{atreya86}, where $v_{\tau}$ is the characteristic velocity of turbulence.  The pressure scale height in the extended thermosphere of HD209458b is $H_p >$~10$^6$ m and consequently it is possible that K$_{\tau} >$~10$^6$~m$^2$~s$^{-1}$.  The coefficients of molecular diffusion reach similar values at $p <$~20~nbar \citep{koskinen10a} indicating that uniform mixing can take place up to the nbar levels.

Drag forces between the escaping hydrogen and minor species can also lead to uniform mixing of the thermosphere.  Whether or not a heavier atom moves along with hydrogen depends on its mass.  If the mass of the atom is smaller than the \textit{crossover mass} $M_c$, the drag forces are strong enough to prevent diffusive separation.  Approximately, the crossover mass is given by \citep{hunten87}: 
\begin{equation}
M_c = M_H + \frac{k T F_H}{b g x_H}
\end{equation}
where $M_H$ is the mass of atomic hydrogen, $F_H$ is the escape flux, and $b$ is the binary diffusion coefficient for hydrogen and the heavier species.  Assuming a mass loss rate is $dM/dt =$~10$^{7}$--10$^{8}$~kg~s$^{-1}$ and a binary diffusion coefficient of $b \sim$~1.6~$\times$~10$^{22}$~m$^{-1}$~s$^{-1}$, we obtain the crossover mass of $M_c =$~11--102.  This implies that drag forces affect the mixing ratios of oxygen and carbon atoms, and could explain a significant abundance of these atoms in the thermosphere.                                              
 
\section{Results}
\label{sc:results}

\subsection{The distribution of hydrogen}
\label{subsc:hydrogen}

In this section we investigate the degree to which the existing UV transit observations of HD209458b are consistent with the simple atmospheric model described in Section~\ref{subsc:model_atmosphere}.  We begin by describing the fitting of the model to the transit depth measurements covering the full width of the stellar H~Ly$\alpha$ emission line profile in the wavelength range of [1212,1220] \AA.  We note that VM04 obtained a line-integrated absorption depth of 5.3~$\pm$~1.8~\% in this wavelength range while BJ10 obtained a higher depth of 6.6~$\pm$~2.3~\%. 

In order to calculate the depth of the transit, we used equation~(\ref{eq:integ_trans}) together with the absorption line parameters listed in Table~\ref{table:spectra}.  We simulated the stellar emissions by using the H~Ly$\alpha$ line profile of HD209458b calculated by \citet{wood05} (hereafter W05).  This line profile is based on a high resolution spectrum obtained with the STIS E140M grating and it was generated by reversing the absorption of the line profile by the ISM.  We fitted a sum of three Gaussian distributions to the pre-ISM line profile to represent the self-reversed core and the broad wings of the line.  We then scaled the line-integrated flux to match with the observations of HD209458b analyzed by BJ07 after accounting for interstellar absorption.         

\begin{deluxetable*}{lccccc}
  \tablecolumns{6}
  \tablecaption{Absorption line parameters$^{a,b}$} 
  \tablehead{
    \colhead{Line}  & 
    \colhead{$\lambda_{0}$(\AA)}  & 
    \colhead{$f_0$} &
    \colhead{$A_0$(s$^{-1}$)} &
    \colhead{$\Delta \lambda_D$(\AA)} &
    \colhead{$\sigma_0$(m$^2$)} 
    }
  \startdata
  HI Ly$\alpha$ & 1215.67  & 4.164~$\times$~10$^{-1}$ & 6.265~$\times$~10$^{8}$ & 0.087 & 5.87~$\times$~10$^{-18}$ \\
  DI Ly$\alpha$ & 1215.34  & 4.165~$\times$~10$^{-1}$ & 6.270~$\times$~10$^{8}$ & 0.062 & 6.23~$\times$~10$^{-18}$ \\
  OI                      &  1302.17 & 4.887~$\times$~10$^{-2}$ & 3.204~$\times$~10$^{8}$ & 0.023 & 2.95~$\times$~10$^{-18}$ \\
  OI*                     &  1304.86 & 4.877~$\times$~10$^{-2}$ & 1.911~$\times$~10$^{8}$ & 0.024 & 2.95~$\times$~10$^{-18}$ \\
  OI**                   &  1306.03 & 4.873~$\times$~10$^{-2}$ & 6.352~$\times$~10$^{7}$ & 0.024 & 2.95~$\times$~10$^{-18}$  
  \enddata
  \tablenotetext{a}{from \citet{morton91} and \citet{morton03}} \\
  \tablenotetext{b}{Doppler FWHMs and central cross sections were calculated for $T =$~10,000~K}
  \label{table:spectra}
\end{deluxetable*}

\begin{figure}
  \epsscale{1.15}
  \plotone{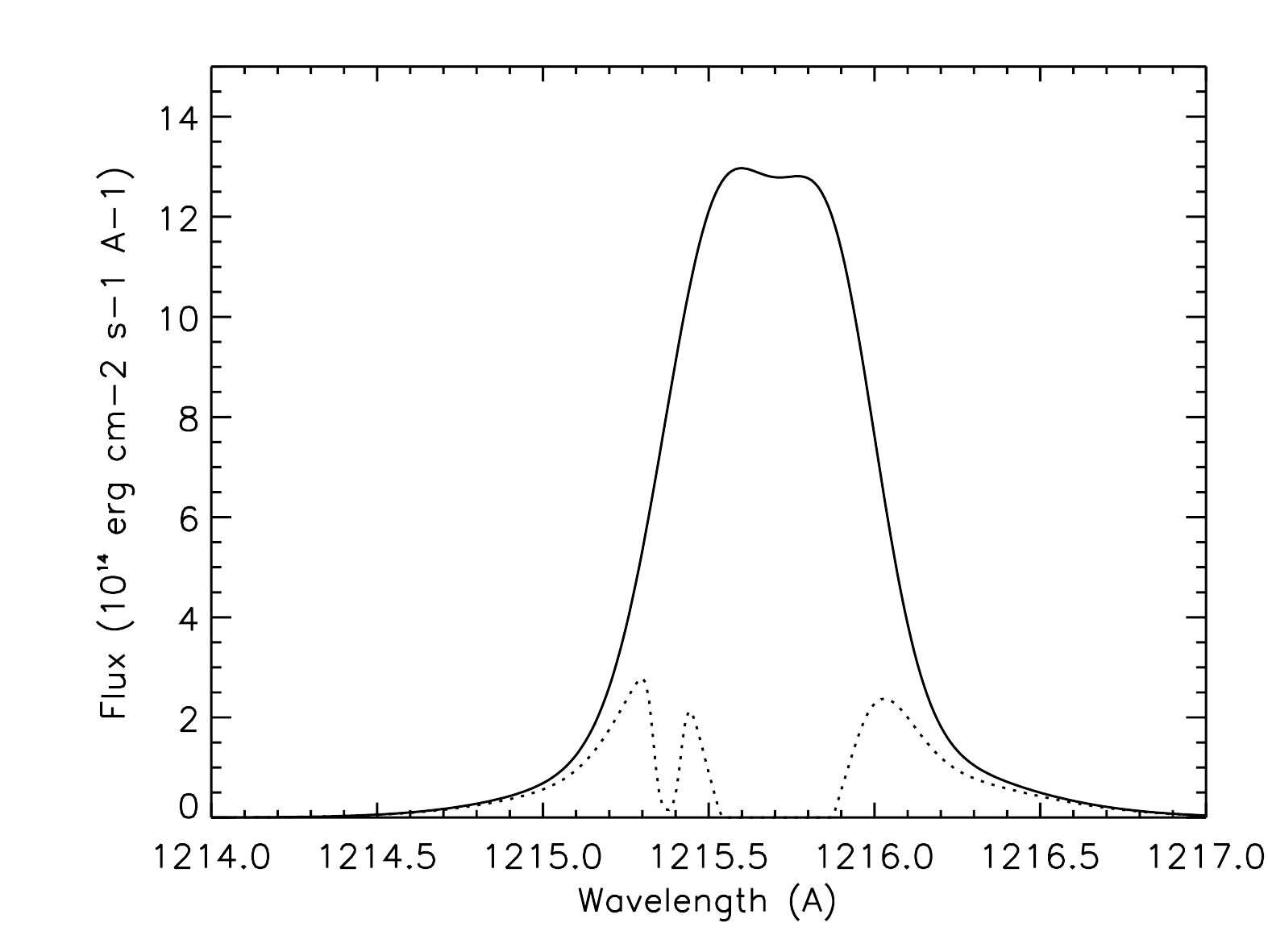}
  \caption{The model H~Ly$\alpha$ profile in the reference frame of the star.  The solid line is the unattenuated out-of-transit stellar flux and the dotted line shows the flux after the absorption by the ISM has been applied.  Two features are visible: a broad absorption band due to interstellar hydrogen (H~I) between 1215.5--1215.9~\AA~and a narrow absorption band due to interstellar deuterium (D~I) centered at a laboratory wavelength of 1215.34~\AA.  Both features are shifted according to the ISM parameters of \citet{wood05}.}
  \label{fig:ism}
\end{figure}

Most of the flux contained within the line profile is absorbed by the ISM, and thus it is important to make sure that the model fluxes are properly attenuated before the line-integrated transit depth is calculated.  W05 measured the column density of hydrogen towards HD209458 and it is $N_H =$~2.3~$\times$~10$^{22}$~m$^{-2}$.  They reported a Doppler broadening parameter along this line of sight of $b_D =$~12~km~s$^{-1}$.  This Doppler parameter is typical of the local interstellar medium (LISM) where the temperature is thought to vary between $T_{ISM} =$~7,000--12,000~K \citep{landsman93}.  The column density, on the other hand, is relatively low and indicates that the sightline to HD209458 is in a low-density direction of the LISM.  We calculated the extinction as a function of wavelength by using these parameters and adopting a Voigt function to simulate thermal and natural broadening in the ISM.  Figure~\ref{fig:ism} illustrates the effect of interstellar absorption on our model line profile.

\begin{figure}
  \epsscale{1.15}
  \plotone{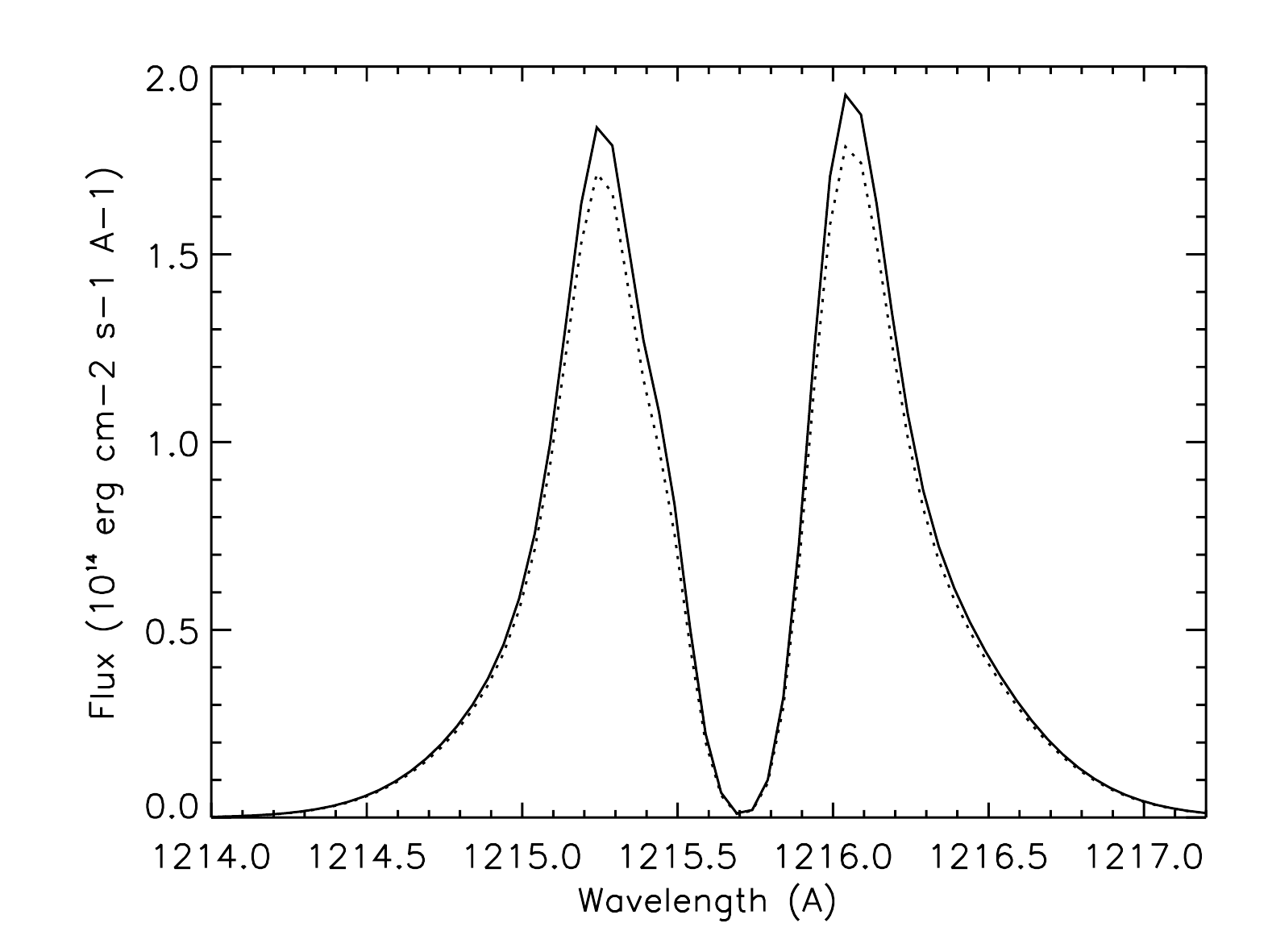}
  \caption{H~Ly$\alpha$ emission lines based on the M1 model (see Table~\ref{table:models}) convolved with an LSF corresponding to the spectral resolution of 0.08~\AA~appropriate for the STIS G140M grating.  The solid line shows the out-of-transit flux and the dotted line shows the in-transit flux.  The full-width transit depth based on these line profiles is 6.6~\%.}
  \label{fig:synthetic_spectrum}
\end{figure}

We obtained the observed full-width H~I transit depth of 6.6~\% by adopting a mean temperature of $T =$~11,000~K and assuming that most of the hydrogen in the atmosphere of HD209458b is ionized above $Z_e \approx$~2.9~R$_p$.  The rest of the free parameter values for this best-fitting empirical model, labeled M1, and the other models discussed in this paper are listed in Table~\ref{table:models}.  In order to illustrate the appearance of the transit  in real observations, we convolved the data with the line-spread function (LSF) appropriate for the STIS G140M grating.  Figure~\ref{fig:synthetic_spectrum} shows the resulting synthetic stellar H~Ly$\alpha$ line in and out of transit.

The location of the upper boundary in the M1 model agrees with models of the thermosphere, which suggest that the H/H$^+$ ionization front is located between $Z_e =$~1.5--5~R$_p$ \citep[][]{yelle04,garciamunoz07}.  It is also close to the boundary of the Roche lobe in the direction perpendicular to the LOS at $Z_{H \perp} \sim$~2.9~R$_p$.  Further, the pressure at $Z_e$ is $p =$~0.04 nbar, which is comparable to the pressure of the solar wind scaled to $a =$~0.047~AU.  Thus, if the planet lacks a significant magnetic field \citep[e.g.][]{griessmeier04}, the upper boundary of the empirical model may be close to the ionopause.  

\begin{figure}
  \epsscale{1.15}
  \plotone{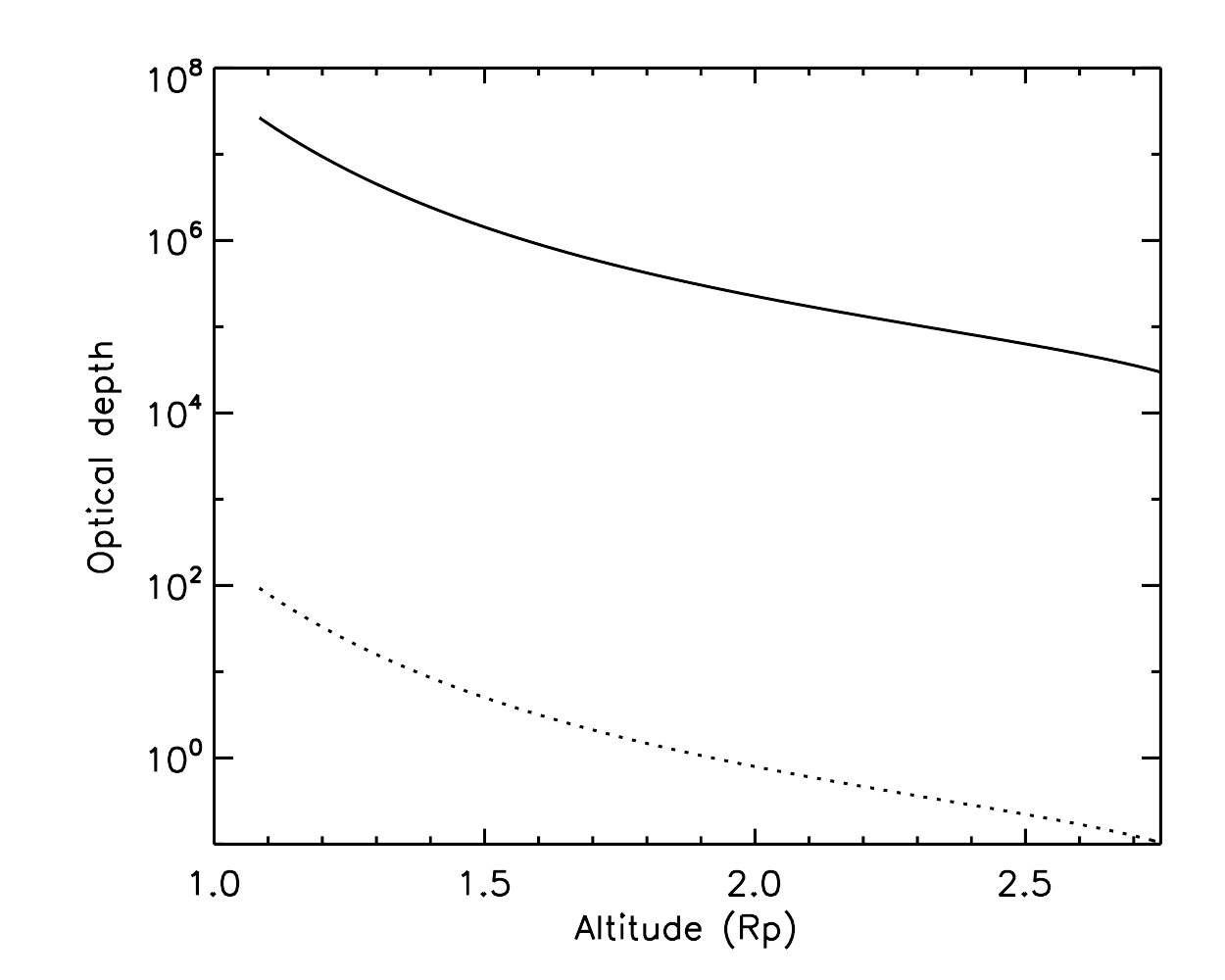}
  \caption{Optical depth of the H~Ly$\alpha$ line as a function of altitude for the best-fitting empirical model atmosphere of HD209458b.  The solid line shows the optical depth at the central wavelength of $\lambda_o =$~1215.67~\AA~and the dotted line shows the optical depth at $\lambda =$~1215.2~\AA.  The latter wavelength corresponds to $\Delta v =$~--116~km~s$^{-1}$ in the stellar reference frame.}
  \label{fig:opdepth}
\end{figure}

The optical depth profiles for the M1 model are shown in Figure~\ref{fig:opdepth} for the core of the line at $\lambda =$~1215.67~\AA~and for the wing of the line at $\lambda =$~1215.2~\AA, which corresponds to $\Delta v \approx$~--116~km~s$^{-1}$ in the stellar reference frame.  We agree with BJ10 that due to thermal and natural broadening, and large column densities along the lines of sight through the atmosphere, the optical depth in the wings of the line profile is significant even in the absence of actual bulk flows towards or away from the observer.  The core of the line is optically thick up to and beyond the upper boundary while the wing is optically thick up to the altitude of $z =$~2~R$_p$.  The latter altitude corresponds to a transit depth of $\sim$5.4~\%.   

\begin{deluxetable*}{ccccccccccc}
  \tablecolumns{9}
  \tablecaption{Model parameters} 
  \tablehead{
    \colhead{ID} &
    \colhead{T~(K)}  & 
    \colhead{Z$_e$~(R$_p$)}  & 
    \colhead{$p_b$~($\mu$bar)} &
    \colhead{$n_b$~(m$^{-3}$)} &
    \colhead{$m_f^a$} & 
    \colhead{$v_r$~(km~s$^{-1}$)} &
    \colhead{$N_{HI}$ (m$^{-2}$)$^b$} &
    \colhead{H~I~(\%)} &
    \colhead{O~I~(\%)} &
    \colhead{$\chi^2$ (H~I)$^c$} 
    }
  \startdata
  M1          & 11,000          & 2.92  & 0.1      & 6.6~$\times$~10$^{16}$ & 1     & 0.0 & 5.8~$\times$~10$^{23}$ & 6.6 & 4.3  & 1.9  \\
  M2          & 11,000          & 2.92  & 0.1      & 6.6~$\times$~10$^{16}$ & 1     & 10  & 5.8~$\times$~10$^{23}$ & 6.6 & 5.8  & 1.9  \\
  M3          & 11,350          & 2.92  & 0.1      & 6.4~$\times$~10$^{16}$ & 5     & 0.0 & 5.7~$\times$~10$^{23}$ & 6.6 & 5.0  & 1.9  \\
  M4          & 12,000          & 2.96  & 0.1      & 6.0~$\times$~10$^{16}$ & 10   & 0.0 & 5.4~$\times$~10$^{23}$ & 6.6 & 5.5  & 1.9  \\
  M5          & 15,950          & 3.08  & 0.1      & 4.5~$\times$~10$^{16}$ & 40   & 0.0 & 4.0~$\times$~10$^{23}$ & 6.6 & 7.1  & 2.2  \\
  M6          & 11,000          & 2.92  & 0.001 & 6.6~$\times$~10$^{14}$  & 1    & 0.0  & 6.1~$\times$~10$^{21}$ & 2.3 & 2.3  & 8.9 \\
  M7          & 8250             & 2.72  & 1         & 8.8~$\times$~10$^{17}$ & 1     & 0.0  & 5.4~$\times$~10$^{24}$ & 6.6 & 3.9  & 1.8 \\
  K10        & 10,280$^d$ & 2.92  & 10       & 6.6~$\times$~10$^{19}$  & 1    & 0.0  & 6.8~$\times$~10$^{23}$ & 5.2 & 3.7  & 3.1 \\
  Y04        & 12,170$^d$ & 3.0     & 200    & 1.9~$\times$~10$^{21}$  & 1    & 2.2$^e$  & 9.9~$\times$~10$^{22}$ & 2.9 & 2.7 & 7.3  \\
  \enddata
  \tablenotetext{a}{Metallicity enhancement.  In practice this factor is used to multiply the O/H and C/H ratios.}
  \tablenotetext{b}{The vertical column density of H~I in the thermosphere.} 
  \tablenotetext{c}{The chi-squared values are provided merely to aid comparison between different models.}
  \tablenotetext{d}{The model includes a self-consistent pressure-temperature profile.  The quoted temperature is an average of the \textit{altitude} levels.}
  \tablenotetext{e}{The stated velocity applies to the upper boundary.} 
  \label{table:models}
\end{deluxetable*}

\begin{figure}
  \epsscale{1.15}
  \plotone{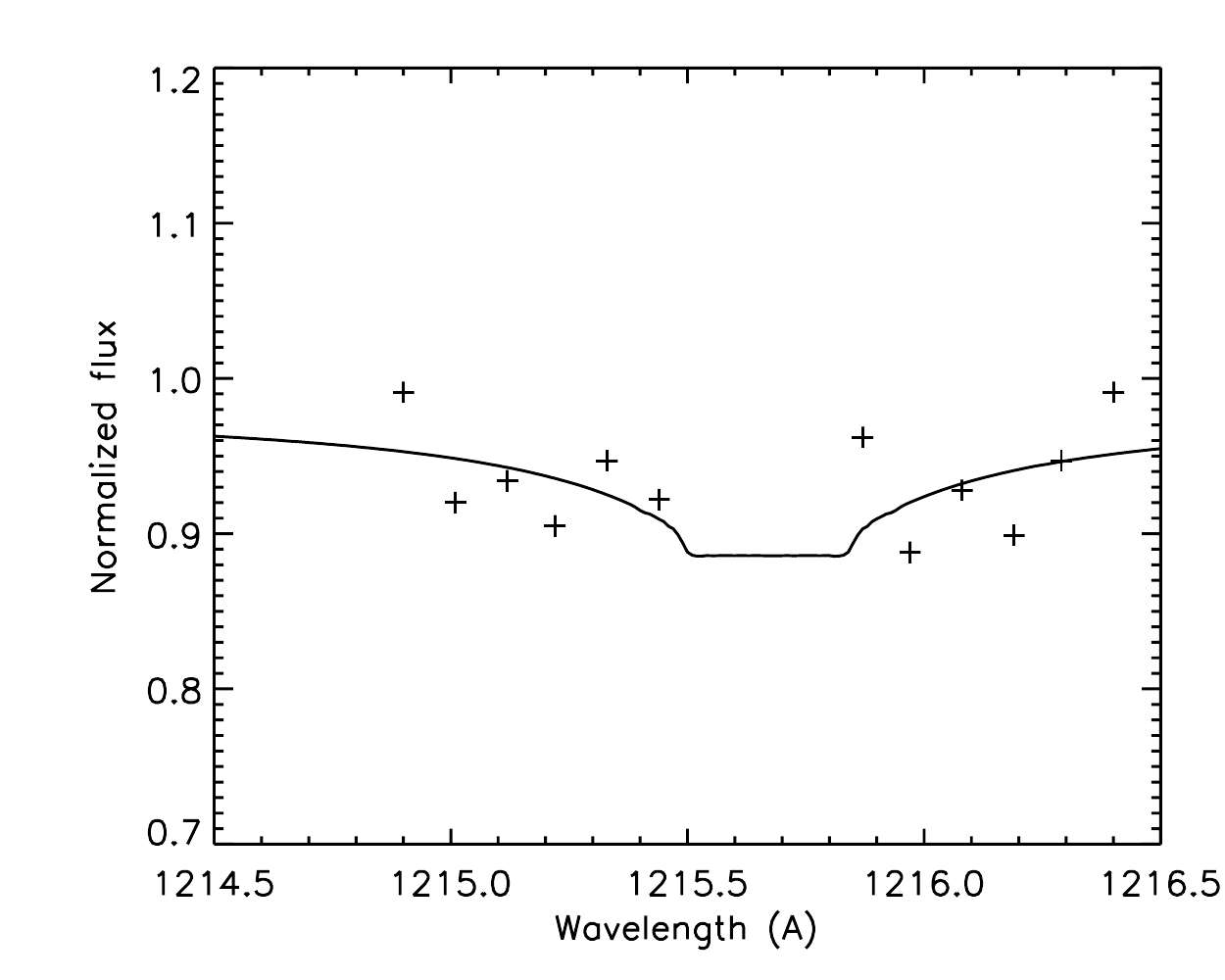}
  \caption{Transmission as a function of wavelength across the stellar H~Ly$\alpha$ emission line during the transit of HD209458b.  The data points were obtained from Figure 6 of BJ08.  The solid line shows our model fit to the data.  The chi-squared value for the fit is $\chi^2 \sim$~1.9.}
   \label{fig:transmission}
\end{figure}

Figure~\ref{fig:transmission} shows transmission of the stellar H~Ly$\alpha$ emission during transit as a function of wavelength based on the M1 model.   It also shows the transmission data points from Figure 6 of BJ08.  The M1 model is consistent with $\sim$11~\% absorption within the core of the line profile.  We calculated $\chi^2 \sim$~1.9 for our fit to the data points.  We note that BJ10 obtained nearly the same value by using the DIV1 model of \citet{garciamunoz07} (hereafter GM07) after multiplying the density profiles of that model by a factor of 2/3.  The results indicate that our fit to the data is as good as that obtained by BJ10.  This is not surprising because in both the M1 empirical model and the DIV1 model the mean temperature is $T \approx$~10,000~K and the density profiles are approximately hydrostatic in the relevant altitude range of $z =$~1.08--3.0~R$_p$.   

\begin{figure}
  \epsscale{1.15}
  \plotone{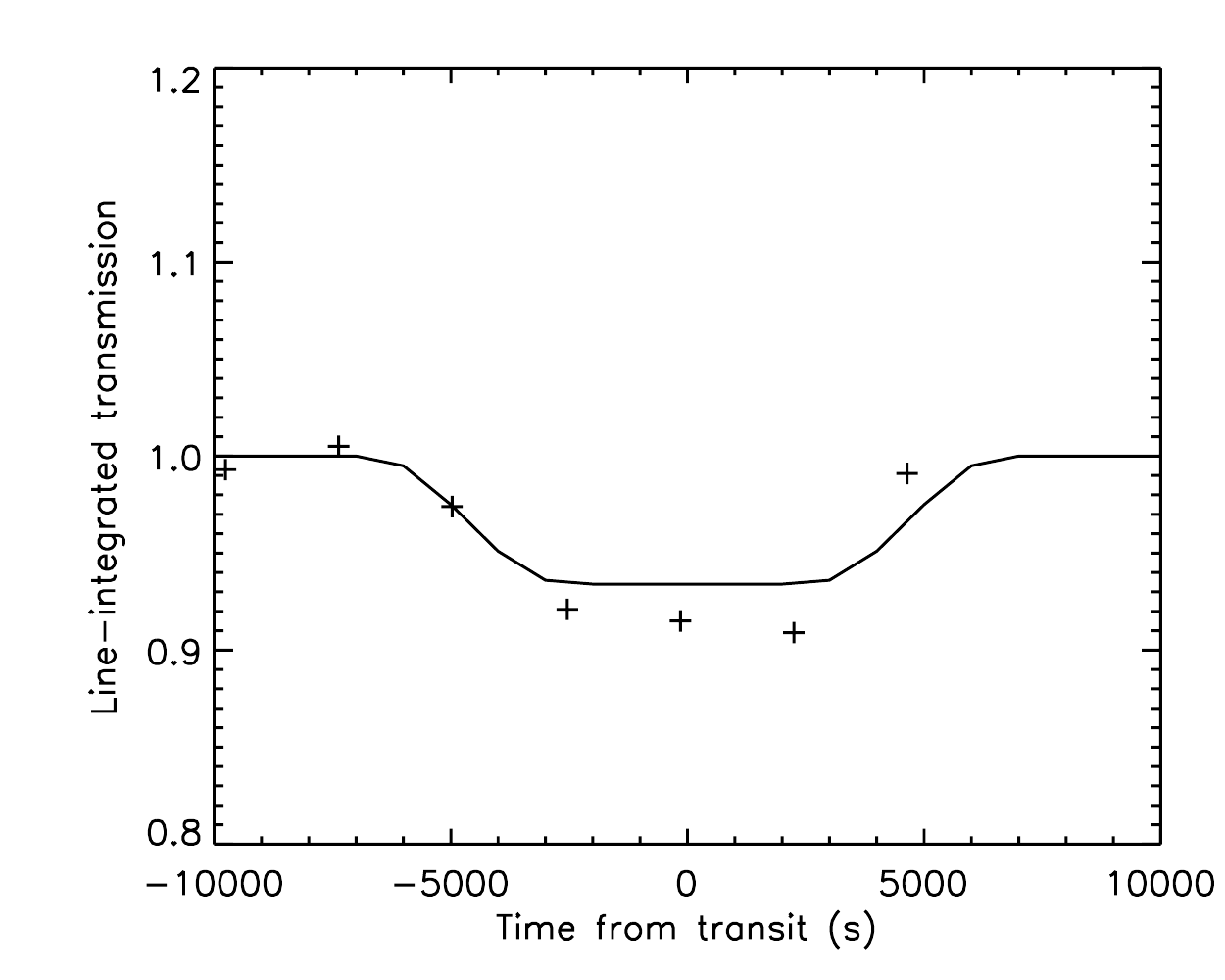}
  \caption{Model light curve showing the H~I transit depth as a function of time from the center of the transit.  The data points were obtained from Figure 2(a) of BJ07 and the solid line is the best-fitting model light curve. The appropriate wavelength limits specified by BJ07 and BJ08 were used to calculate the transit depths.} 
   \label{fig:light_curve}
\end{figure}

Figure~\ref{fig:light_curve} shows the M1 model light curve as a function of time from the center of the transit compared with the observed data points from Figure 2(a) of BJ07.  In order to calculate this light curve, we used the wavelength limits specified by BJ07 and BJ08.  The transit depth at the center of the transit is $\sim$6.6~\%, which is slightly lower than that given by BJ07 but the difference is not statistically significant.  The symmetry of the light curve and a good fit by our model indicate that there is no observational evidence that the extended thermosphere of HD209458b deviates significantly from a spherical shape.  This is in contrast to the findings of VM03 who, based on the asymmetry of their light curve, argued that the planet is followed by a curved cometary tail of escaping hydrogen.

Next we explore whether the full width transit depth of 5--7~\% is consistent with the larger depth of 8--15~\% measured over a more limited wavelength range within the line profile (VM03,BJ07,BJ08).  VM03 obtained the depth of 15~$\pm$~4~\% by comparing the ratio of the integrated flux within the wavelength intervals of [1215.15,1215.5] \AA~and [1215.8,1216.1] \AA~(the blue and red sides of the In region, respectively) to the flux within the intervals of [1214.4,1215.15] \AA~and [1216.1,1216.8] \AA~(the blue and red sides of the Out region, respectively) obtained during the transit.  In their analysis, the interval [1215.5,1215.8] \AA~was excluded because contamination by the geocorona and sky background makes the flux measurements in this region unreliable.  BJ07 redefined all of these wavelength intervals and obtained the transit depth of only 8.9~$\pm$~2.1~\%.  In his analysis the `contamination window' was extended to [1215.41,1215.84] \AA~and the blue and red intervals were defined as [1214.83,1215.36] \AA~and [1215.89,1216.43] \AA, respectively.  

Another significant difference between the analysis of VM03 and BJ07 is the treatment of the time variability of the H~Ly$\alpha$ emissions, either due to stellar variability or noise.  BJ07 derived unperturbed H~Ly$\alpha$ profiles by merging the subspectra of a dense time series that was corrected for the transit trend.  In the resulting line profiles (Figure 3 in BJ07), the time variability has been reduced to $\sim$1~\% level.  VM03, on the other hand, attempted to account for time variability by calculating the transit depth based on the core-to-wing flux ratio instead of evaluating the transit depth based on the ratio of the flux obtained during transit to the flux obtained outside of the transit.  This approach relies on the assumption that the transit depth in the wings of the line is not significant and that the variability in the wings is similar to the variability within the core of the line.                  

\citet{vidalmadjar08} (hereafter VM08) correctly pointed out that a comparison between the two approaches is not fair because BJ07 used different wavelength limits and instead of evaluating the change in the core-to-wing (In/Out) flux ratio, he calculated the true transit depth i.e., the flux decrement based on equation~(\ref{eq:integ_trans}).  However, we believe that the transit depths calculated by VM03 are not necessarily related to the size of the absorbing cloud while those calculated by BJ07 and BJ08 are.  By using the M1 model and the VM03 and BJ07 wavelength limits, we obtain absorption depths of 2.4~\% and 1.3~\%, respectively, based on the In/Out flux ratio and true transit depths of 7.5~\% and 6.6~\%, respectively.  Our model is a good fit to the transmission curve of BJ08 and the behavior of the In/Out ratio is roughly in line with the line profiles shown in Figure 3(a) of BJ07 and yet the transit depth based on the In/Out flux ratio is small.  We note, however, that the M1 model overflows the Roche lobe and thus we agree with VM08 that the atmosphere is evaporating.

\subsection{The distribution of oxygen}
\label{subsc:oxygen}

The large transit depths in the O~I and C~II lines reported by VM04 and BJ10 can potentially be used to constrain the degree of eddy mixing and the composition of the upper atmosphere.  The O~I transit depth, if accurately measured, is a good indicator of the proportion and mixing of heavy elements in the thermosphere.  The interpretation of the C~II transit depth, on the other hand, is much more complicated because of the uncertainties related to the ionization profiles, photochemistry and, if C~II is escaping the thermosphere, the interaction of the atmosphere with the stellar wind or magnetospheric plasmas.  Statistically, the observed C~II transit depth is also the least significant detection out of the three currently available UV transit depth measurements.  For these reasons, we limit our discussion and analysis to the interpretation of the O~I transit depth.            

VM04 reported an integrated transit depth of 12.8~$\pm$~4.5~\% in the wavelength range of [1300,1310]~\AA, which covers the O~I triplet emission lines listed in Table~\ref{table:spectra}.  BJ10 reported a revised and slightly shallower transit depth of 10.5~$\pm$~4.4~\%.  These transit depths are only 2.4 and 1.9~$\sigma$, respectively, away from the FUV continuum transit depth of $\sim$2~\%.  The data were obtained by using the STIS G140L grating with a spectral resolution of $\sim$2.5~\AA, which does not allow for the individual lines of the triplet to be resolved.  The lines were resolved, however, in an earlier spectrum obtained by VM03 with the E140M echelle grating.  Although this spectrum was not used to analyze the transit depths, it is a good reference point for the expected line profiles and can be used to constrain the properties of the ISM.  

We modeled the emission spectrum of the O~I triplet by using the full disk solar line profile parameters derived by \citet{gladstone92} and fitted the line-integrated fluxes to the observations.  As a result we obtained surface fluxes of 3.0, 3.2, and 3.4 W~m$^{-2}$ for the O~I, O~I*, and O~I** lines of HD209458, respectively.  In order to correct the model fluxes for absorption by interstellar oxygen, we assumed that the O~I/H~I ratio in the LISM is $\sim$3~$\times$~10$^{-4}$ \citep{moos02} and that all of the interstellar oxygen is in its ground state.  The latter assumption is consistent with the ratios of the individual emission lines in the high-resolution E140M spectrum.  

BJ10 used the E140M data points to deduce fluxes for the O~I, O~I*, and O~I** lines that scale as 1:1.5:1.17.  The solar line-integrated fluxes of these lines scale as 1:1.06:1.16.  After applying interstellar absorption to the solar line profiles, we obtained model fluxes that scale as 1:1.43:1.55.  The close agreement between the model and observed O~I/O~I* flux ratios confirms that most of the interstellar oxygen is in its ground state and that our parameters for the LISM are reliable.  The observed O~I**/O~I flux ratio, on the other hand, is slightly lower than the model flux ratio.  This cannot be due to absorption in the ISM because a population inversion in the excited levels is unlikely.  Thus the disagreement, if real, arises from the intrinsic properties of HD209458 and implies that the stellar O~I triplet flux ratios differ slightly from the corresponding solar flux ratios.  We note, however, that the E140M spectrum is noisy and it is not clear if the difference is real or if it arises from uncertainties in the data.          

The flux in the core of the O~I ground state line is absorbed by the ISM, and a significant O~I transit depth is possible only if the populations of the excited states of oxygen in the thermosphere of HD209458b are sufficient.  We estimated the relative population of the ground state and the excited states by assuming that the thermosphere is in LTE and by using the electronic partition function calculated by \citet{colonna09}.  At a temperature of $T =$~10,000~K we obtained population ratios of 5.2:3.0:1 for the ground state, lower excited level, and upper excited level, respectively.  These ratios agree closely with the statistical weights of the levels that give rise to the triplet.   

The M1 model presented in Table~\ref{table:models} yields an O~I transit depth of 4.3~\%.  This is similar to the transit depth calculated by BJ10 who used the scaled density profiles from the DIV1 model of GM07.  Again, the agreement between these models is not surprising because the temperature and density profiles are similar and both models assume solar O/H ratios.  We note that the O~I transit depth based on the M1 model is within $\sim$1.4$\sigma$ from the observed depth.  Assuming Gaussian statistics, the probability that the true transit depth is between 1.7--6.1~\% (2--1$\sigma$) is $\sim$14~\%.  Due to the relatively large uncertainty in the data points, this probability is significant.  In other words, the deviation of the model transit depth from the observed value is not statistically significant.  Better observations with higher S/N and spectral resolution are required to constrain the transit depth further.  Our analysis shows that the STIS G140M grating, rather than the G140L grating, could already be used to obtain better data.          

BJ10 argued that the O~I transit depth of 10.5~\% is possible if oxygen is preferentially heated to T$_{OI} \sim$~100,000~K (8.6~eV) within a layer located between $z =$~2.25~R$_p$ and $Z_{H \perp} =$~2.9~R$_p$.  It is true that hot oxygen populations have been observed in the thermospheres of terrestrial planets \citep[e.g.][]{cotton93,shematovich06,nagy81} but these populations are formed mostly by dissociative recombination of O$_{2}^+$, which should not be important on HD209458b.  Other possible mechanisms for heating oxygen atoms include collisional processes that, as we demonstrate in the next section, would also imply that oxygen is mostly ionized.  Thus it is not clear if a neutral cloud at these temperatures is physically possible.  

As we already pointed out, the O~I transit depth of 4.3~\% based on the M1 model is consistent with the observations.  However, the measurements also allow for a larger transit depth and it is thus worth exploring if transit depths close to 10.5~\% can be explained by changing the parameters of the model atmosphere, the host star, or the interstellar medium.  In the following we investigate several different possibilities that can give rise to larger O~I transit depths without the need to introduce energetic oxygen atoms in the thermosphere.       

Theoretically, the O~I transit depth is sensitive to interstellar absorption.  In order to test if changes in the abundance of interstellar oxygen can increase the calculated transit depth, we assumed that the column density of interstellar oxygen towards HD209458b is negligible.  As a result, the calculated transit depth increased only by fractions of a percent.  We conclude that, contrary to H~I,  the full-width O~I transit depth of HD209458b is not very sensitive to interstellar absorption.  In addition, the E140M spectra indicate that our ISM parameters are accurate and the assumption of a negligible oxygen column would not be reasonable anyway.  Consequently, we must explore the properties of the host star and the planet in order to explain the possibility of large O~I transit depths.          

Transit depths in the visible and the IR wavelengths are sensitive to stellar activity and can appear larger if a significant portion of the stellar disk is covered by starspots -- even if these spots do not actually fall in the path of the transit \citep[e.g.,][]{czesla09}.  This is because the effective emitting area of the star appears smaller.  In order to crudely estimate the magnitude of this effect, we defined an \textit{effective} fractional area, $A_q$, of the quiet regions as the ratio of the area that would be required to emit zero flux to account for the diminished luminosity during quiet periods to the area of the stellar disk.  In terms of $A_q$, the observed transit depth, $d_T$, can be written as:   
\begin{equation}
d_T \approx \frac{\left( R_p / R_s \right)^2}{1 - A_q}
\end{equation}
where $R_p$ is the radius of the transiting planet, and $R_s$ is the radius of the star.  We note that $A_q$ would have to be $\sim$60~\% for the O~I transit depth to increase from 4.3~\% to 10.5~\%.  Because the definition of $A_q$ implies that the \textit{actual} fractional area of the quiet regions would have to be much larger than 60~\%, this is clearly not realistic.  In fact, the O~I triplet flux of the Sun varies by $\sim$30~\% during the solar cycle \citep{gladstone92}, and thus we would expect an enhancement at most by 0.7 percentage points (even this is probably overestimated because the relationship between the flux variability and $A_q$ is complicated).  This is not statistically significant and we conclude that the interpretation of the low-resolution transit depths is not affected by the distribution of active or quiet regions.  
     
The transit depth measurements are also affected by stellar limb darkening or limb brightening.  According to the solar disk observations of the O~I triplet fluxes reported by \citet{rousseldupre85}, the fluxes are roughly constant between $\mu (= \cos \theta) \sim$~1--0.2, with only negligible limb darkening.  At $\mu <$~0.2 the fluxes are brighter than within the center of the disk by a factor of $\sim$1.1.  If the atmosphere of the transiting planet does not cover the bright ring surrounding the stellar disk, limb brightening actually makes the transit depth appear shallower.  We simulated limb brightening by dividing the stellar disk into a central disk region and a bright ring, and neglected limb darkening within the central region.  As a result, we found that limb brightening does not lead to a detectable change in the low-resolution O~I transit depth.     

Several authors have attempted to explain the observed O~I transit depth by referring to the velocity dispersion arising from radial outflows.  VM04 invoked this explanation and argued that hydrodynamically escaping oxygen atoms moving at sonic or supersonic velocities could give rise to a transit depth of 13~\%.  Later, \citet{tian05} argued that the velocity dispersion arising from similar supersonic outflows could also explain the large H~I transit depth reported by VM03.  They assumed that the Doppler broadening parameter is given by:
\begin{equation}
\Delta \nu_D = \frac{\nu_o}{c} \sqrt{ \frac{2 k T}{m} + \eta^2 }
\label{eqn:broadening}
\end{equation}
where $\eta$ is the velocity dispersion. Assuming that $\eta =$~10~km~s$^{-1}$, and using the M1 model as a base, we obtain H~I and O~I transit depths of 6.7 and 7.7~\%, respectively.  Chi-squared for the H~I transmission fit in this case is $\chi^2 \sim$~2.1, which indicates that the H~I transit depth is not significantly affected by additional velocity dispersion.  This is because a large portion of the H~Ly$\alpha$ line is already saturated.  On surface, then, velocity dispersion appears to be a promising way to explain the large O~I and C~II transit depths.     

Equation~(\ref{eqn:broadening}), however, is appropriate for simulating broadening due to \textit{microturbulence}.  It is only accurate if the scale of the turbulence is much smaller than the mean free path.  Further, what really matters is not radial velocity but the projected velocity along the LOS.  Because most of the absorption during transit is due to the layers near the terminator, radial velocity is not a good measure of the true velocity dispersion.  In order to estimate the velocity dispersion properly, we calculated the appropriate velocities for each layer along each LOS and used these velocities in integrating equation~(\ref{eq:tau}).  For each layer $k$ along a LOS, the parallel velocity $v_{k \parallel}$ is given by:   
\begin{equation}
v_{k \parallel} (z) = v_r (r_k) \cos \lbrack \frac{\pi}{2} - \cos^{-1} \left( \frac{z}{r_k} \right) \rbrack
\end{equation}
where $v_r$ is the radial velocity, and $z$ is the constant tangent altitude of the LOS.  Assuming a uniform radial velocity of $v_r \sim$~10~km~s$^{-1}$ for the M1 model, the maximum LOS velocity, obtained at the upper boundary of the model along a sightline through the lower boundary is $v_{\infty \parallel} (z_b) \sim \pm$9.2~km~s$^{-1}$.  For a LOS at $z =$~2~R$_p$, the corresponding velocity is only $v_{\infty \parallel}(2~R_p) = \pm$6.8~km~s$^{-1}$.

\begin{figure}
  \epsscale{1.15}
  \plotone{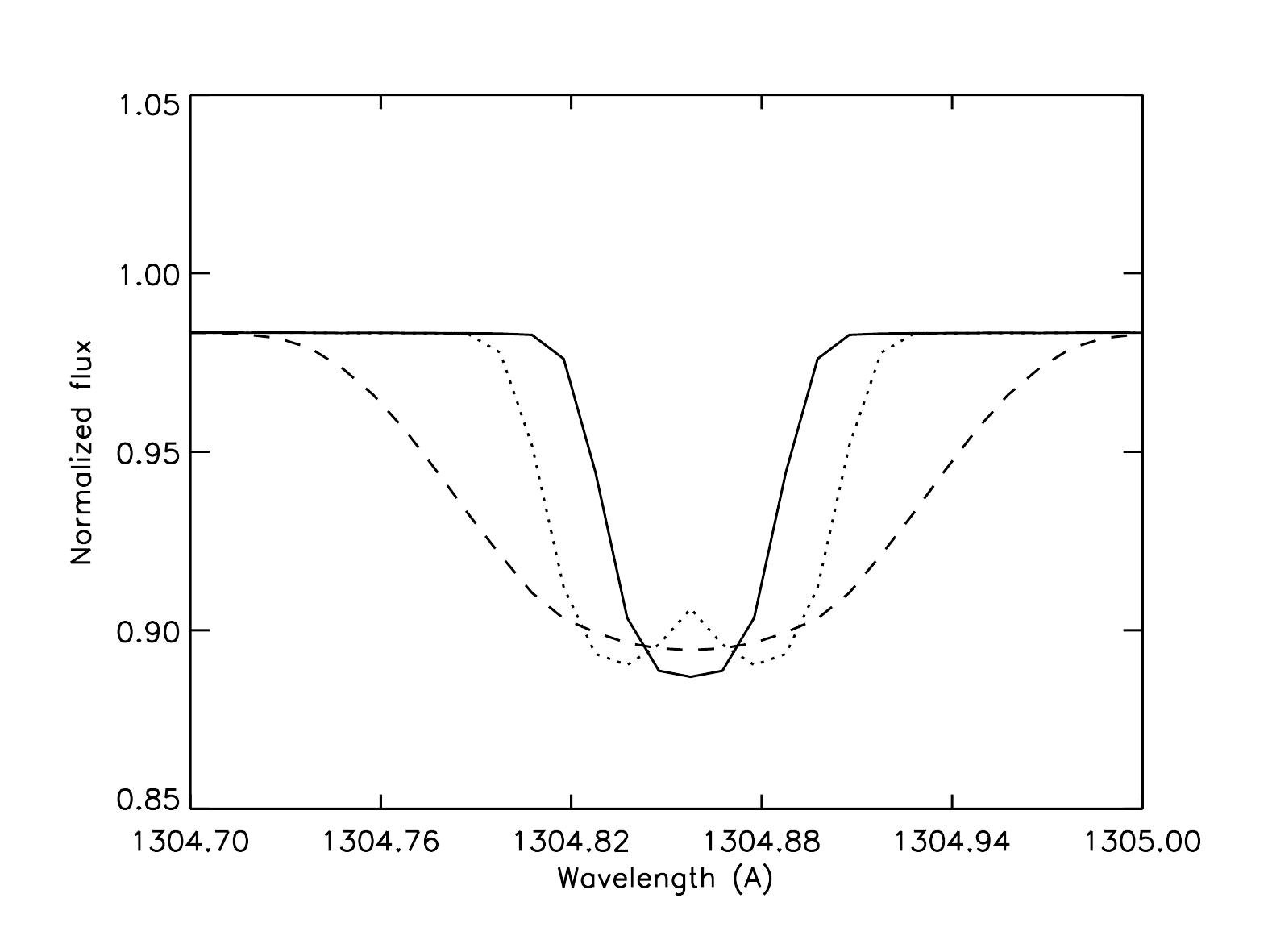}
  \caption{Transmission as a function of wavelength in the O~I* line.  The solid and dotted lines show transmission based on the M1 and M2 models, respectively (see Table~\ref{table:models}).  The M2 model includes a uniform radial velocity of $v_r =$~10~km~s$^{-1}$ and thus the absorption line profile exhibits two symmetric peaks around the central wavelength of $\lambda_o =$~1304.86~\AA.  The dashed line shows transmission for a model that includes the same radial velocity dispersion but by using equation~(\ref{eqn:broadening}) for microturbulence.}   
    \label{fig:profiles}
\end{figure}

The M2 model (see Table~\ref{table:models}) includes a uniform radial velocity of $v_r =$~10~km~s$^{-1}$.  Figure~\ref{fig:profiles} shows the transmission profile of the O~I* line for this model together with a corresponding profile for the M1 model and another profile calculated by using equation~(\ref{eqn:broadening}).  The absorption profiles differ significantly.  By using equation~(\ref{eqn:broadening}) we obtain a much broader absorption profile with a slightly smaller central cross section compared to the M1 model.  If the velocity dispersion is treated properly, as it is in the M2 model, the absorption profile exhibits two symmetric peaks around the central wavelength of the O~I* line at $\lambda_o =$~1304.86~\AA.  The resulting H~I and O~I transit depths based on the M2 model are 6.6 and 5.8~\%, respectively.

A radial velocity of $v_r =$~10~km~s$^{-1}$ is not enough to explain an O~I transit depth of 10.5~\% but it can explain O~I transit depths that are comparable to the H~I transit depth even when the abundance of oxygen is significantly lower than that of hydrogen.  It is not realistic, however, to assume that $v_r =$~10~km~s$^{-1}$ in the lower thermosphere.  Hydrodynamic models of the thermosphere indicate that the base of the planetary wind is located near $p =$~1~nbar, which corresponds to an altitude of $z =$~1.7~R$_p$ in the M1 model.  Assuming that the outflow is effective above this altitude, we obtain an O~I transit depth of only 5.5~\%.  Consequently, velocity dispersion can only affect the transit depth significantly if the optical depth above the critical level of $z_c =$~3~R$_p$ is significant.

It should be noted that radial velocities higher than $v_r =$~10~km~s$^{-1}$ do not necessarily lead to larger transit depth because the absorption is limited by the fact that the central cross section in equation~(\ref{eq:sigma0}) is proportional to $\Delta \nu_D$.  In fact, if we introduce a radial velocity of $v_r =$~60~km~s$^{-1}$ to the M1 model, we obtain an O~I transit depth of only 5.2~\%.  Within the velocity range of $v_r =$~10--60~km~s$^{-1}$, we obtain a maximum O~I transit depth of 7.2~\% by assuming that $v_r =$~30~km~s$^{-1}$.  Thus fast hydrodynamic escape of neutral oxygen from the thermosphere alone cannot explain an O~I transit depth of 10.5~\%. 

The next logical step is to investigate if the O/H and C/H ratios in the thermosphere of the planet are actually higher than the solar values.  BJ10 did not consider this possibility because the density profiles of their best-fitting model are already enhanced with respect to the Y04 model.  With respect to the DIV1 model of GM07, however, their density profiles are actually depleted and further, the O/H and C/H ratios of their model are not enhanced but solar.  We generated new empirical models M3, M4, and M5 (see Table~\ref{table:models}) by increasing the O/H and C/H ratios of the M1 model by factors of 5, 10, and 40, respectively.  Initially, this caused both the H~I and O~I transit depths to decrease because the higher mean molecular weight of the thermosphere led to a shorter pressure scale height.  In fact, we found that it is not possible to increase the abundance of the heavier atoms without also increasing the mean temperature of thermosphere to recover the fit to the H~I transit depth.  In order to do so, we increased the mean temperatures of the M4 and M5 models to $T \approx$~12,000~K and 16,000~K, respectively.  The M3 model did not require a significantly increased temperature.  

The mean temperature of the thermosphere can be constrained by two additional means.  First, the temperature of the neutral atoms is limited by thermal ionization.  In LTE conditions the thermal ionization balance is determined by the Saha equation, which implies that atomic hydrogen is fully ionized at temperatures higher than $T =$~7000--8000~K.  In non-LTE conditions of the thermosphere the ion/neutral fractions resulting from thermal ionization are determined by a balance between impact ionization by thermalized electrons and radiative recombination.  The reaction rate coefficients of these processes, respectively, for atomic hydrogen are \citep{voronov97,storey95}:
\begin{eqnarray}
k &=& 2.91~\times~10^{-14} \frac{U^{0.39}}{U + 0.232} \exp( -U ) \  \   m^3~s^{-1} \\
& & \nonumber \\
&\alpha_A& = 4.2~\times~10^{-19} ( 10^4 / T_e )^{0.64} \  \   m^3~s^{-1} 
\end{eqnarray}
where $T_e$ is the electron temperature and $U = I_p / E_e$ is the ratio of the ionization potential $I_p$ to the energy of the electrons $E_e \approx k T_e$.  The balance of these rate coefficients implies that $\lbrack$H$^+ \rbrack$ / $\lbrack$H$\rbrack  >$~100 for $T_e >$~25,000~K.  Based on the impact ionization rate of \citet{voronov97} and the radiative recombination rate of \citet{schunk00}, the ionization fraction of atomic oxygen i.e., $\lbrack$O$^+ \rbrack$ / $\lbrack$O$\rbrack  >$~100 in the same temperature range.

Second, as BJ10 also points out, higher temperatures lead to a worsening fit to the H~I transmission data in Figure~\ref{fig:transmission}.  As the temperature approaches $T =$~15,000~K, $\chi^2 >$~2 for the fit.  This is not a very strong constraint at all, of course, but it indicates a general trend that disfavors temperatures much higher than $T =$~15,000~K.  Indirectly, the need to retain the fit to the H~I transmission data and these temperature limits imply that models with O/H and C/H ratios much higher than 40 times solar ratios are not realistic, and that enhanced oxygen abundance can increase the O~I transit depth at most to 7.1~\%.  We also experimented with a combination of supersolar oxygen abundance and hydrodynamic escape by introducing a uniform radial velocity of $v_r =$~30~km~s$^{-1}$ to the M5 model.  As a result, we obtained an O~I transit depth of 10.8~\%, which is actually close to 10.5~\%.  Despite being a good fit to the observations, there are several good reasons for why this model is unrealistic.  The O/H ratio of the M5 model is 40 times higher than the solar ratio and $v_r =$~30~km~s$^{-1}$ below the critical level.  Matching the observations is not enough prove a theory -- the theory also has to make physical sense.       

It is possible that, contrary to our assumptions, there are optically thick clouds of neutral hydrogen and oxygen above the upper boundary of the M1 model moving at large LOS velocities.  These clouds can be formed, for instance, by neutral atoms that escape the atmosphere and react with the stellar wind \citep{holstrom08,ekenback10}.  Although the density profile of the empirical model is misleading at high altitudes, we extended the M1 model thermosphere to $z = $~10~R$_p$.  As a result, we obtained H~I and O~I transit depths of 13.1 and 12.2~\%, respectively.  In this case, $\chi^2 =$~13.4 for the H~I transmission fit.  In a slightly more realistic case, we assumed that only 10~\% of the atoms are neutral between $Z_e$ and $z = $~10~R$_p$.  This led to H~I and O~I transit depths of 7.8 and 6.3~\%, respectively, and $\chi^2 =$~2.1.  We note that external neutral atoms can affect the transit depth but we also agree with BJ10 that the H~I transmission data points are not consistent with optically thick populations of hydrogen moving at large velocities.  Detailed models of the chemical and dynamical interaction of the escaping atmosphere with the stellar wind outside the Roche lobe are required to study the distribution of external neutral atoms.                  

\subsection{Models for the thermosphere of HD209458b}
\label{subsc:models}

With the exception of GM07, most of the modeling efforts so far have concentrated on simulating the distribution of hydrogen (and sometimes helium) in the escaping atmosphere of HD209458b.  These models are justified by the assumption that the \textit{overall} characteristics of the thermosphere are not affected by small abundances of heavier neutral atoms or ions.  In addition to the empirical model presented in Section~\ref{subsc:model_atmosphere}, we have used two models with different characteristics and limiting assumptions to calculate transit depths.  These models are the one-dimensional outflow model of \citet{yelle04} (Y04) and a thermospheric circulation model developed by \citet{koskinen10a} (K10).  The photochemistry in these models is based on a composition of hydrogen and helium only.  In order to roughly estimate the O~I transit depth, we added a solar proportion of oxygen to these models and assumed an oxygen ionization fraction equal to that of hydrogen.    

The K10a model is different from the existing outflow models in that it is three-dimensional and can account for horizontal flows that are important in the thermosphere of HD209458b.  On the other hand, the model assumes that the thermosphere is in hydrostatic equilibrium and thus it cannot properly account for vertical flows that lead to a deviation from hydrostatic conditions.  Also, the photochemical scheme of the model is not valid in regions where H$^+$ is the dominant species.  To account for these limitations, we placed the upper boundary of the transit depth calculations to $z =$~2.9~R$_p$ in line with the arguments we have made for the validity of the empirical model.  In order to calculate the optical depth as a function of altitude, we averaged the density and temperature profiles of the model globally to obtain one-dimensional vertical profiles.  The resulting H~I and O~I transit depths are 5.2 and 3.7~\%, respectively, and for the H~I transmission fit we obtain $\chi^2 \sim$~3.1 (see Table~\ref{table:models}).  These values are in rough agreement with our earlier calculations and the observations.        

The one-dimensional outflow model developed by Y04 solves the vertical component of the Navier-Stokes equations and includes a self-consistent treatment of much of the important physics in the thermosphere of HD209458b.  The altitude range of the model is limited to $z <$~3~R$_p$ because above this altitude stellar gravity, radiation pressure, and interaction of the escaping atmosphere with the stellar wind affect the solution in ways that cannot be accounted for by simple one-dimensional, hydrodynamic models.  In order to ensure consistency with kinetic theory, the velocity at the upper boundary of the model was required to be equal to the weighted mean of the velocity distribution functions of different species.  This is not the same as setting the velocity to be equal to Jeans effusion velocity because \textit{drifting} Maxwellians were used to calculate the boundary condition in an iterative fashion.      

The density and temperature profiles of the Y04 model yield H~I and O~I transit depths of 2.9 and 2.7~\%, respectively (see Table~\ref{table:models}).  While the full-width H~I transit depth is within 1.6$\sigma$ of the observed value, $\chi^2$ for the HI transmission curve is larger than the one obtained for the empirical model or the K10a model.  The reason for the slightly poorer fit lies in the details of the density profiles.  The H$_2$/H dissociation front of the Y04 model is located at the pressure of $p =$~0.5~nbar at the altitude of $z =$~1.1~R$_p$.  This means that the model is less extended than the empirical model below the 0.5~nbar level because the mean molecular weight of the gas is larger and the thermosphere is cooled effectively by infrared emissions from H$_{3}^{+}$ ions.  Also, the ionization front of the model is located at a relatively low altitude of $Z_e \approx$~1.7~R$_p$.  

The extent and temperature of the thermosphere depend strongly on the location of the H$_2$/H dissociation front that effectively determines the number density and temperature at the lower boundary of the empirical model.  In order to illustrate this point, we changed the pressure at the lower boundary of the M1 model from $p_b =$~0.1~$\mu$bar to $p_b =$~1~nbar (see model M6 in Table~\ref{table:models}).  As a result we obtained reduced H~I and O~I transit depths of 2.3 and 2.2~\%, respectively.  This explains why \citet{murrayclay09} obtained a relatively low H~I transit depth of $\sim$3~\% by using a one-dimensional outflow model with a composition based on H and H$^+$ only.  The pressure and temperature at the lower boundary of the model are only $p \approx$~30~nbar and $T_b =$~1000~K, respectively.  Thus the model ignores the critical region of the atmosphere that contributes significantly to the optical depth.  If the pressure of their lower boundary was higher, they could have obtained a larger transit depth.         

In the K10a model the H$_2$/H dissociation front is located at the pressure of $p =$~0.1~$\mu$bar due to a combination of thermal dissociation and horizontal mixing.  In the DIV1 model of GM07 it is located at $p =$~1.5~$\mu$bar because, contrary to the other models, it includes ion and neutral photochemistry with carbon, oxygen, and nitrogen species.  Thus H$_2$ is also dissociated in reactions with atomic oxygen and the OH radical.  We note, however, that \citet{liang03} also modeled the neutral photochemistry and found that H$_2$ was not entirely converted into H at these pressure levels.  Motivated by the DIV1 model, we moved the lower boundary of the M1 model to $p_b =$~1~$\mu$bar and obtained H~I and O~I transit depths of 6.6 and 3.9~\%, respectively (see model M7 in Table~\ref{table:models}).  In this case, we adjusted the mean temperature of the thermosphere to $T =$~8250~K in order to retain the fit to the H~I transit depth.  In terms of optical depth, the M7 model is almost indistinguishable from the M1 model.    

Our exploration together with the results of BJ10 reveal that models of the thermosphere can be used to match the observed H~I absorption depths as long as the location of the H$_2$/H dissociation front, the temperature and the location of the upper boundary agree with the values derived from our empirical model.  These findings provide strong justification for our approach of using a simple empirical model to fit the data.  We would not have been able to identify these important parameters that affect the optical depth profiles in the thermosphere if we had allowed one of the more complicated models to cloud our judgement.  Once known, these parameters can help to guide the hydrodynamic and photochemical models towards solutions that agree with the observed transit depths.                                           

\section{Discussion}
\label{sc:discussion}       

In this section we assess the feasibility of our basic assumptions and results.  We find that it is easy to generate models that agree with the observed H~I transit depths.  According to the best-fitting empirical models M1 and M7 (see Table~\ref{table:models}), the H$_2$/H dissociation front is located between $p =$~0.1--1~$\mu$bar ($z \approx$~1.1~R$_p$) and the mean temperature of the thermosphere is between $T =$~8000--11,000~K.  We note that due to thermal ionization, the temperature of the neutral thermosphere cannot be much higher than  $T \approx$~15,000~K.  The M1 model atmosphere is optically thick up to $z \approx$~2~R$_p$ in the wings of the H~Ly$\alpha$ line and at least up to the Roche lobe in the core of the line.  Above our upper boundary at $Z_e =$~2.9~R$_p$, however, the atmosphere is mostly ionized.   

Our results are sensitive to the pressure level of the H$_2$/H dissociation front.  Additional observations are required to constrain the vertical column density of H~I in the thermosphere.  Such observations were recently obtained by \citet{france10} who attempted to observe the auroral and dayglow emissions of H$_2$ from HD209458b.  They did not detect the emissions and argued that this implies that the vertical column density of H~I must be at least $N_{HI} =$~3~$\times$~10$^{24}$~m$^{-2}$ to prevent stellar FUV radiation from exciting H$_2$ in the lower thermosphere.  Table~\ref{table:models} includes substellar column densities of H~I for the models discussed in this paper.  According to this additional constraint, it is possible that the H$_2$/H dissociation front is deeper than the 0.1~$\mu$bar level.  In that case, M7 is the best-fitting model.       

The detection or non-detection of infrared emissions from H$_{3}^{+}$ can also be used to constrain the properties of the thermosphere.  \citet{shkolnik06} attempted to observe these emissions from a sample that included the transiting planet GJ436b and several non-transiting EGPs by using the CSHELL spectrograph at the NASA Infrared Telescope Facility (NIRTF).  They reported upper limits for the fluxes that are often several orders of magnitude higher than the predicted fluxes for close-in EGPs.  More recently, \citet{swain10} obtained a near-IR secondary eclipse spectrum of HD189733b by using the SpeX instrument at NIRTF.  The lack of a 3.9~$\mu$m feature in this spectrum may imply that H$_{3}^{+}$ emissions are not important on this particular planet.  We note that a lack of emissions implies that there is not enough H$_2$ to form H$_{3}^{+}$ or that H$_{3}^{+}$ is destroyed in reactions with oxygen or carbon species.  

Modeling the chemistry and dynamics of the lower thermosphere is a complicated problem.  For instance, GM07 has shown that the complex photochemistry that includes oxygen and carbon species affects the location of the H$_2$/H dissociation front and thus the extent and temperature of the thermosphere.  H$_2$ can also be dissociated by high-energy electrons that penetrate below the 1~$\mu$bar level and have not been included in the existing photochemical models.  The composition of the atmosphere is also affected by dynamics.  Because the upper atmosphere is strongly ionized, the conductivities are likely to be high.  As a result, the dynamics of the thermosphere is affected by strong electric currents that have not been included in the existing models either.          

The results are not as sensitive to the location of the upper boundary as they are to the location of the lower boundary.  We note, however, that the location of the upper boundary at $Z_e =$~2.7--2.9~R$_p$ in both our work and that of BJ10 is remarkably consistent.  It is tempting to suggest that this is not a coincidence.  After all, the upper boundary is close to both the boundary of the Roche lobe and the ionopause of a weakly magnetized planet.  It is possible that, for some reason, the density of the neutral atoms decreases sharply above this altitude.  We believe that multidimensional plasma models, either kinetic or MHD, are needed to study the density profiles of different species above the upper boundary.     

Based on an earlier study by \citet{holstrom08}, \citet{ekenback10} (hereafter E10) studied the formation of energetic neutral atoms (ENAs) outside the Roche lobe of HD209458b.  They pointed out that the ENAs are created by charge exchange between hydrogen atoms escaping the atmosphere and the protons of the stellar wind.  They argued that absorption by ENAs can explain the H~I transmission curve reported by VM03.  In principle, this is a good idea but some of the details are unclear.  We are not convinced that the boundary conditions of the E10 model are consistent with realistic models of the thermosphere.  Also, the E10 model was not used to fit the BJ08 transmission data.    

The inner boundary of the E10 model is located at $z =$~2.8~R$_p$ and the optical depth of the \textit{atmosphere} in the H~Ly$\alpha$ line is assumed to be small inside the inner boundary sphere.  At the same time the number density at the inner boundary is $n =$~4~$\times$~10$^{13}$~m$^{-3}$ because a mass loss rate of $dM/dt \sim$~1.6~$\times$~10$^{11}$~g~s$^{-1}$ is required to ensure a sufficient production rate of the ENAs.  We note that the number density of the M1 model at $z =$~2.9~R$_p$ is $n =$~2.6~$\times$~10$^{13}$~m$^{-3}$ and that the density profile of this model already leads to a full-width H~I transit depth of 6.6~\%.  Thus absorption by ENAs is not required to fit the data.  On the contrary, too much additional absorption may actually lead to a deteriorating fit to the detailed H~I transmission data.

The mass loss rate of $dM/dt \sim$~1.6~$\times$~10$^{11}$~g~s$^{-1}$ is in line with the GM07 model.  BJ10 showed that this model can be used to fully explain the observed absorption.  They also pointed out that an optically thick external population of neutral hydrogen is possible but the LOS velocities must be slow.  In this case, additional absorption is constrained to the core of the line, which is fully absorbed by the ISM and the transmission in the wings is not significantly affected.  Our analysis supports this conclusion.  New models that treat the thermosphere and the external region self-consistently are required to study the true distribution of external neutral atoms.  

The O~I transit depth is less precise than the H~I transit depth but we are able to explain the observations with our empirical model and O/H ratios ranging from solar to 40 times solar, although the latter value implies an uncomfortably high temperature.  The large O~I transit depth indicates that the thermosphere is uniformly mixed up to the Roche lobe although there is a marginal possibility that this is not the case because the FUV continuum transit depth of 2~\% is within the 1.96$\sigma$ confidence interval of the detected depth.  If diffusive separation takes place, the O~I transit depth cannot be much higher than the continuum transit depth.  

Based on the M1 model, we obtain a full-width O~I transit depth of 4.3~\%, which is statistically consistent with the observed absorption.  In this case, oxygen overflows the Roche lobe with hydrogen.  Assuming radial velocities of $v_r =$1--10~km~s$^{-1}$ at $Z_e =$~2.9~R$_p$, the total mass loss rate from the M1 model is $dM/dt \sim$~4--40~$\times$~10$^{7}$~kg~s$^{-1}$ and this leads to a total escape flux of $F_{O} =$~1.8--18~$\times$~10$^{31}$ s$^{-1}$ for oxygen.  We have also shown that the velocity dispersion due to radial outflow, supersolar oxygen abundance and external populations of neutral oxygen can explain O~I transit depths that are comparable to the full-width H~I depth.  We have not, however, been able to identify \textit{feasible} mechanisms that give rise to O~I transit depths that are significantly higher than the H~I depth.          

We believe that more precise measurements are required to further constrain the O~I transit depth.  We have generated synthetic data by simulating the response function of the STIS G140M grating to show that, by assuming the M1 model to be accurate, we would observe a transit depth of 4.3~$\pm$~1~\% for HD209458b.  Thus for transit depths higher than $\sim$5~\%, a confident detection in the upper atmosphere can easily be obtained.  In addition, the spectral resolution of the G140M grating is high enough to resolve the individual emission lines of the O~I triplet.  As we have seen, spectrally resolved measurements provide much better constraints on the actual transit depth than unresolved observations.  We believe that it would be useful to repeat the O~I transit depth measurements by using this instrument.             

\section{Conclusion}
\label{sc:conclusion}

Transmission spectroscopy in UV wavelength bands is a rich source of information about the upper atmospheres of extrasolar planets.  In this paper we have introduced a generic method that can be used to interpret and analyze UV transit light curves.  This method is based on tracing the emitted flux from the stellar atmosphere through the atmosphere of the planet and the interstellar medium to the observing instrument at Earth.  We have demonstrated the method in practice by applying it to the existing UV transit depth measurements of HD209458b in the H~Ly$\alpha$ and O~I triplet lines (VM03,VM04,BJ10).  In order to interpret the measured transit depths we used a simple empirical model of the thermosphere based on the generic features of more complex models \citep[e.g.,][]{yelle04,garciamunoz07,koskinen10a} to simulate absorption by the occulting atmosphere.        

The H~I transit depth is sensitive to interstellar absorption and the full-width transit depth reflects the optical depth of the atmosphere in the wings of the H~Ly$\alpha$ line.  We found that it is easy to generate models that explain the observed absorption without the need to introduce external ENAs or hydrogen atoms accelerated by radiation pressure outside the Roche lobe.  According to the best-fitting models (see models M1 and M7 in Table~\ref{table:models}), the mean temperature of the thermosphere is $T =$~8000--11,000~K and the H$_2$/H dissociation front is located at $p_b =$~0.1--1~$\mu$bar ($z_b \approx$~1.1~R$_p$).  The upper boundary of the model is located at $Z_e \sim$~2.9~R$_p$, which is near the boundary of the Roche lobe.  Below $Z_e$ the density profiles are approximately hydrostatic and above $Z_e$ the atmosphere is mostly ionized.  

By using the M1 model, we obtain a full-width H~I transit depth of 6.6~\% and the model transmission of the H~Ly$\alpha$ line matches the data points reported by BJ08.  The apparent disagreement between BJ08 and the earlier analysis by VM03 arises from differences in the treatment of the data and different definitions of the transit depth.  In particular, there is no definite observational evidence for a cometary tail following the planet in the transit light curve and the absorption is not significantly stronger in the blue side of the H~Ly$\alpha$ line.  However, we agree with VM08 that the core of the absorption line is optically thick up to the Roche lobe and thus that the atmosphere is evaporating.  We estimate a mass loss rate of $dM/dt \approx$10$^{7}$--10$^{8}$~kg~s$^{-1}$ based on the density profile of the M1 model and a range of possible radial velocities.          

It is possible that ENAs are present outside the Roche lobe as suggested by E10, but the optical depth of the ENA clouds has to be consistent with the underlying models of the thermosphere.  We do not agree with E10 that the optical depth of the thermosphere below $z =$~2.8~R$_p$ is small.  We do believe, however, that one-dimensional hydrodynamic models are not adequate in modeling the distribution of ionized gases outside the Roche lobe.  Multidimensional plasma models are required to study the interaction of the ionosphere with the stellar wind and magnetospheric plasma self-consistently.      

The mean O~I transit depth of 10.5~\% is only 1.93$\sigma$ away from the FUV continuum transit depth of $\sim$2~\%.  By using the M1 model, we obtain an O~I transit depth of 4.3~\%, which is marginally consistent with the observations.  Full-width O~I transit depths that are comparable or slightly higher than the H~I depth are possible if the abundance of oxygen in the thermosphere is supersolar, the atmosphere is escaping with supersonic velocities or if large external clouds of neutral oxygen are present above the Roche lobe.  More precise measurements are required to constrain the O~I transit depth further.  Out of the currently available instruments, the STIS G140M grating can be used to obtain repeated measurements with higher S/N and spectral resolution.                            

\acknowledgments

The authors would like to thank B. E. Wood for correspondence about the H~Ly$\alpha$ emission line profile of HD209458b.  T.T.K. would like to thank J. Y-K. Cho for helpful discussions regarding the structure and dynamics of the ionosphere.  This research has been supported by the NASA's Planetary Atmospheres Program through grant NNX09AB58G.

\end{document}